\RequirePackage{fix-cm}
\documentclass{svjour3}                     

\usepackage{nicefrac}
\usepackage[hyphens]{url}
\usepackage{soul}
\usepackage{hyperref}
\hypersetup{breaklinks=true}
\usepackage[utf8x]{inputenc}
\usepackage{xspace}
\usepackage{makecell}
\usepackage{xcolor}
\usepackage[position=bottom]{subfig}
\usepackage{float}
\usepackage{graphics}
\usepackage{graphicx}
\usepackage{color, colortbl}
\usepackage{caption}
\usepackage{wrapfig}
\usepackage{paralist}
\usepackage{booktabs} 
\usepackage{array}
\usepackage{balance}
\usepackage{lipsum}
\usepackage{optidef} 
\usepackage{dsfont}  
\usepackage[ruled,norelsize]{algorithm2e}
\usepackage{algorithmic}
\makeatletter
\newcommand{\removelatexerror}{\let\@latex@error\@gobble}
\makeatother

\usepackage{amsmath}

\usepackage{tikz}
\usepackage{physics}
\usepackage{cancel}
\usepackage{siunitx}
\usepackage{multirow}
\usepackage{tabularx}

\usetikzlibrary{arrows.meta}
\usetikzlibrary{positioning}
\usetikzlibrary{fadings}
\usetikzlibrary{patterns}
\usetikzlibrary{shadows.blur}
\usetikzlibrary{shapes}
\usetikzlibrary{shapes.symbols} 

\usepackage{adjustbox}
\usepackage{bbm}

\usepackage{mathptmx}   

\usepackage{appendix}
\tikzset{
  >={Latex[width=2mm,length=2mm]},
            base/.style = {rectangle, rounded corners, draw=black, fill=gray!10,
                           minimum width=2.3cm, minimum height=1cm,
                           text centered, font=\large\sffamily},
            data/.style = {rectangle, draw=black, fill=gray!10, minimum width=2.3cm, minimum height=1cm, text centered, node distance=1cm and 1cm, font=\large\sffamily},
            corpus/.style={tape, tape bend top=none,
                draw=black, fill=gray!10, minimum width=2.3cm, minimum height=1cm, text centered, font=\large\sffamily},
            bias/.style= {draw=none, 
                          fill=none,
                          minimum width=1.1cm,
                          text=red,
                   font=\bf\large\sffamily},
            plaintext/.style= {draw=none, 
                          fill=none,
                          minimum width=1.1cm,
                          text=black,
                 font=\bf\large\sffamily},
            process/.style= {base,
                             node distance=1cm and 1cm,
                             fill=gray!25, font=\large\sffamily},
            node/.style = {base, 
                           node distance=1cm and 1cm, font=\large\sffamily}
}

\smartqed    

\newcommand*{\eg}{e.g.,\xspace}
\newcommand*{\ie}{i.e.,\xspace}
\newcommand*{\etc}{etc.\xspace}

\newcommand*{\etal}{\emph{et~al.}\xspace}
\newcommand*{\humantic}{\texttt{Humantic AI}\xspace}
\newcommand*{\crystal}{\texttt{Crystal}\xspace}

\newcommand{\rev}[1]{\textcolor{black}{#1}}
\newcommand{\revv}[1]{\textcolor{black}{#1}}

\newcommand*{\signifhigh}{\cellcolor[HTML]{fff719}}
\newcommand*{\signifmedhigh}{\cellcolor[HTML]{fff719}}
\newcommand*{\signifmed}{\cellcolor[HTML]{fff719}}
\newcommand*{\signifmedlow}{\cellcolor[HTML]{fff719}}
\newcommand*{\signiflow}{\cellcolor[HTML]{fffc99}}

\usepackage{amssymb}
\usepackage{pifont}
\newcommand{\cmark}{\ding{51}}%
\newcommand{\xmark}{\ding{55}}
\newcommand{\qmark}{\texttt{?}}

\usepackage{listings}
\definecolor{darkgray}{rgb}{0.33, 0.33, 0.33}

\lstnewenvironment{Python}[1][]
  {\lstset{language=Python,
    basicstyle      = \scriptsize\ttfamily,
    keywordstyle    = \color{blue},
    keywordstyle    = [2] \color{teal}, 
    stringstyle     = \color{magenta},
    literate={ü}{{\"u}}1 {ö}{{\"o}}1 {É}{{\'E}}1 {œ}{{\oe}}1,
    commentstyle    = \color{darkgray}\ttfamily,
           morekeywords=as,
           morekeywords=with,
           #1}%
  }
  {}

\journalname{Data Mining and Knowledge Discovery Special Issue on Bias in Fairness in AI}

\begin{document}
\title{An External Stability Audit Framework to Test the Validity of Personality Prediction in AI Hiring}

\author{Alene K. Rhea       \and
        Kelsey Markey       \and
        Lauren D'Arinzo     \and
        Hilke Schellmann \and 
        Mona Sloane         \and
        Paul Squires        \and
        Falaah Arif Khan    \and 
        Julia Stoyanovich
}

\institute{Alene K. Rhea \at
              Center for Data Science, New York University \\
              Center for Responsible AI, Tandon School of Engineering, New York University \\
              \email{alene@nyu.edu}            
           \and
           Kelsey Markey \at
              Center for Data Science, New York University \\
              Center for Responsible AI, Tandon School of Engineering, New York University \\
              \email{kelseymarkey@nyu.edu}
               \and
           Lauren D'Arinzo \at
              Center for Data Science, New York University \\
              Center for Responsible AI, Tandon School of Engineering, New York University\\
              The MITRE Corporation, Bedford, MA\\
              The author’s affiliation with The MITRE Corporation is provided for identification purposes only, and is not intended to convey or imply MITRE's concurrence with, or support for, the positions, opinions, or viewpoints expressed by the author.Approved for Public Release; Distribution Unlimited. Public Release Case Number 21-3193.© 2021 The MITRE Corporation. All rights reserved.\\
              \email{lauren.darinzo@nyu.edu}
              \and
           Hilke Schellmann \at 
           Arthur L. Carter Journalism Institute, New York University \\
           \email{hilke.schellmann@nyu.edu}
           \and
           Mona Sloane \at 
           Center for Responsible AI, Tandon School of Engineering, New York University \\
           \email{mona.sloane@nyu.edu}
           \and
           Paul Squires \at 
           Department of Psychology, Arts \& Science, New York University \\
           \email{ps2937@nyu.edu}
           \and
           Falaah Arif Khan \at
              Center for Data Science, New York University \\
              Center for Responsible AI, Tandon School of Engineering, New York University \\
              \email{fa2161@nyu.edu}
               \and
           Julia Stoyanovich \at 
           Center for Responsible AI, Tandon School of Engineering\\
           Computer Science \& Engineering,  Tandon School of Engineering\\
           Center for Data Science, New York University \\
           \email{stoyanovich@nyu.edu}
}

\maketitle

\begin{abstract}

Automated hiring systems are among the fastest-developing of all high-stakes AI systems.  Among these are algorithmic personality tests that use insights from psychometric testing, and promise to surface personality traits indicative of future success based on job seekers' resumes or social media profiles.  We interrogate the validity of such systems using stability of the outputs they produce, noting that reliability is a necessary, but not a sufficient, condition for validity. 
Crucially, rather than challenging or affirming the assumptions made in psychometric testing --- that personality is a meaningful and measurable construct, and that personality traits are indicative of future success on the job --- we frame our \revv{audit} methodology around testing the underlying assumptions made by the vendors of the algorithmic personality tests themselves. 

\revv{Our main contribution is the development of a socio-technical framework for auditing the stability of algorithmic systems. This contribution is supplemented with an open-source software library that implements the technical components of the audit, and can be used to conduct similar stability audits of algorithmic systems. We instantiate our framework with the audit of two real-world personality prediction systems, namely Humantic AI and Crystal. The application of our audit framework demonstrates that both these systems show substantial instability with respect to key facets of measurement, and hence cannot be considered valid testing instruments.} 
\end{abstract}


\section{Introduction}
\label{sec:intro}
AI-based automated hiring systems \rev{are seeing ever broader use and have become} as varied as \rev{the traditional hiring practices they augment or replace}.  \rev{These systems} include candidate sourcing and resume screening to help employers identify promising applicants, video and voice analysis to facilitate the interview process, and algorithmic personality assessments that purport to surface personality traits indicative of future success.   HireVue, a company that sells one of these systems, estimates that the ``pre-hire assessment'' market is worth \$3 billion annually~\cite{kelly-lyth_challenging_2020}. Indeed, most Fortune 500 companies are using some form of algorithmic hiring~\cite{pod_ep1}. Ian Siegel, the CEO of ZipRecruiter (a popular online employment marketplace), estimates that 75\%-100\% of all submitted resumes are now read by software, and that only a small fraction of those go on to be read by humans~\cite{pod_ep1}. 

In this paper, we focus on automated pre-hire assessment systems, as some of the fastest-developing of all high-stakes uses of AI~\cite{kelly-lyth_challenging_2020}.  The popularity of automated hiring systems in general, and of pre-hire assessment in particular, is due in no small part to the hiring sector's collective quest for efficiency.  Employers choose to use them to source and screen candidates faster and with less paperwork and, in a world reshaped by the COVID-19 pandemic, with as little in-person contact as is practical.  Job seekers are, in turn, promised a more streamlined job search experience, although they rarely have a choice in whether they are screened by an automated system, and they are typically not notified when algorithmic screening is used~\cite{WSJ_Stoyanovich}). The \rev{flip side} of efficiency potentially afforded by automation is that job seekers, the general public, and even employers themselves rarely understand how these systems work and, indeed, whether they work. Is a resume screener identifying promising candidates or is it picking up irrelevant --- or even discriminatory --- patterns from historical data, potentially exposing the employer to legal liability?  {\it Are  job seekers participating in a fair competition if they are systematically unable to pass an online personality test, despite being well-qualified for the job}~\cite{WSJ_KyleBehm}? 

Personnel selection is an especially sensitive, high-stakes application of AI. Hiring decisions are often of great consequence to candidates' financial and emotional well-being~\cite{bendick_situation_2007}, and in aggregate contribute to widespread economic inequality~\cite{blau_trends_2013,hegewisch_separate_2010}. Consequences for hiring organizations can be substantial as well: if their selection procedures are arbitrary or unfair, they risk litigation and class action lawsuits. As such, any algorithms deployed in the field of hiring deserve rigorous scrutiny.  

\revv{This realization is starting to be codified in laws and regulation.  An important recent example is Local Law 144 of 2021 that requires bias auditing of ``automated employment decision tools'' used by employers in New York City, and also mandates disclosure about the use of these tools to job seekers before they are screened~\cite{nyc144}.  Another example is the Artificial Intelligence Act (AI Act), proposed by the European Commission in 2021 to serve as a common regulatory and legal framework for AI in the European Union~\cite{euaiact}. The Act states that ``AI systems used in employment, workers management and access to self-employment, notably for the recruitment and selection of persons, for making decisions on promotion and termination and for task allocation, monitoring or evaluation of persons in work-related contractual relationships, should also be classified as high-risk, since those systems may appreciably impact future career prospects and livelihoods of these persons,'' and subjects such systems to strict oversight requirements.}

Reports of algorithmic hiring systems acting in ways that are discriminatory or unreliable abound~\cite{bandy_problematic_2021,bogen_help_2018,amazon_bias,datta_automated_2015,friedman_bias_1996,kochling2020discriminated,stark_physiognomic_2021}.  In a recent example, when testing automated phone interview software, Hilke Schellmann found that the system produced ``English competency'' scores even when the candidate spoke exclusively in German or Chinese~\cite{pod_ep2}. This finding undermines the \emph{validity} of the tool, and crystallizes the fact that black-box algorithms may not act as we expect them to.

In our work we interrogate the validity of algorithmic pre-hiring assessment systems of a particular kind: those that purport to estimate a job seeker's personality based on their resume or social media profile.  Our focus on these systems is warranted both because the science behind personality testing (algorithmic or not) in hiring is controversial~\cite{emre_personality_2018,lussier_temperamental_2018,mona_blog}, and because algorithmic personality tests are rarely validated by third-parties~\cite{pod_ep1}. Warning against this trend, Chamorro-Premuzic \etal~\cite{chamorro-premuzic_new_2016} write in the Journal of Industrial and Organizational Psychology: ``shiny new talent identification objects often bamboozle recruiters and talent acquisition professionals with no regard for predictive validity.'' Despite this warning,  unvalidated use of these ``objects'' continues. For example, as we will discuss in Section~\ref{sec:audit}, DiSC, a psychometric instrument used by several algorithmic personality assessment systems, has not been validated in the hiring domain, and the company that produces DiSC specifically warns against using it for pre-employment screening. 

\revv{In our work, we focus on \emph{stability}, by which we refer to a property of an algorithmic system whereby small changes in the input lead to small changes in the output, noting that this property is a necessary, albeit not a sufficient, condition for validity.  Our approach is to (1) develop a methodology for an \emph{external audit of the stability} of algorithmic personality predictors, and (2) instantiate this methodology in an audit of two real-world systems, \humantic and \crystal.  Crucially, based on the insights of Sloane \etal~\cite{sloane_silicon_2021}, we frame our methodology around \emph{testing the underlying assumptions made by the vendors of the algorithmic personality tests themselves.}}

\humantic and \crystal were selected as audit subjects because they each produce quantitative personality traits as output, accept easily-manipulated textual features as input, and allow multiple input types. These systems also have substantial presence in the algorithmic hiring market: \humantic reports that it is used by Apple, PayPal and McKinsey,\footnote{\url{https://humantic.ai/}} and \crystal claims that 90\% of Fortune 500 companies use their products\rev{, though neither company distinguishes between use for hiring and use for other purposes, such as sales}.\footnote{\url{https://www.crystalknows.com/}} 
    
\begin{figure}
\centering 
\includegraphics[width=0.75\textwidth]{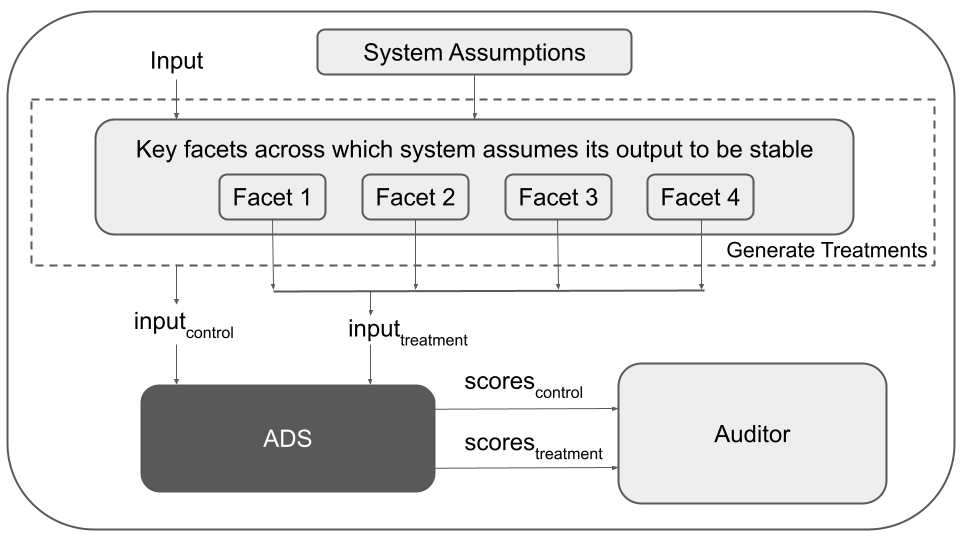}
\caption{\revv{Socio-technical framework for stability auditing, discussed in detail in Section~\ref{sec:methodology}.}}
\label{fig:audit_diagram}
\end{figure}

\paragraph{In this paper, we make the following contributions:} 

\begin{enumerate}
    \item We provide an overview of the key literature on psychometric testing applied to hiring and on algorithm auditing with a particular focus on hiring (Section~\ref{sec:background}). We find that reliability is seen as a crucial aspect of the validity of a psychometric instrument, yet it has not received substantial treatment in algorithm audits. 
    
    \item  \revv{We propose a socio-technical framework for auditing the stability of algorithmic systems (Section~\ref{sec:methodology}).  Figure~\ref{fig:audit_diagram} gives an overview of our proposed methodology. As part of this contribution, we develop an open-source software library that implements the technical components of the audit, and can be used in stability audits of automated decision systems (ADS), with suitable input data, treatment generation techniques, and choice of stability metrics.  Our library can be extended with additional input and output data types, treatment and control generation methods, and choice of stability metrics.  }
    
    \item \revv{We instantiate this methodology in an external stability audit of \humantic and \crystal, two black-box algorithms that predict personality for use in hiring, over a dataset of job applicant profiles collected through an IRB-approved study (Section~\ref{sec:audit})}.  
    
    \revv{The application of our audit framework surfaces substantial instability with respect to important facets of measurement in both these systems, results are presented in Section~\ref{sec:results}.  For example, we find that personality profiles returned by both \humantic and \crystal are substantially different depending on whether they were computed based on a resume or a LinkedIn profile, violating the assumption that an algorithmic personality test is stable across input sources that are treated as interchangeable by the vendor.  Further, \crystal frequently computes different personality scores if the same resume is given in PDF vs. in raw text format, violating the assumption that the output of an algorithmic personality test is stable across job-irrelevant variations in the input.} 
\end{enumerate}

We discuss the results and limitation of our work in Section~\ref{sec:conc}, and conclude in Section~\ref{sec:outlook}.
\section{Background and Related Work}
\label{sec:background}

\subsection{Validity and Reliability in Psychometric Theory Applied to Hiring}
\label{sec:psychometric_theory}

\paragraph{Personality testing in hiring.} Since the early 1900s, personnel selection practices have relied on the use of psychometric instruments such as personality tests to identify promising candidates~\cite{scroggins_psychological_2008}, and the use of these tests continues to be wide-spread~\cite{HR_survey}.  And although this practice is both longstanding and wide-spread, it has been met with skepticism from industrial-organizational (I-O) psychologists due to validity and reliability concerns, and even led to disagreements about whether personality itself is a meaningful and measurable construct~\cite{scroggins_psychological_2008}.  A comprehensive literature review of personality testing in personnel selection published in 1965 found little evidence of predictive validity, and concluded that ``it is difficult to advocate, with a clear conscience, the use of personality measures in most situations as a basis for making employment decisions''~\cite{guion_validity_1965}. Several other surveys would come to the same conclusion in the following decades~\cite{hough_criterion-related_1990,schmitt_metaanalyses_1984}, yet, HR professionals continued to use personality testing for hiring~\cite{scroggins_psychological_2008}. The rise of the ``Big Five'' model of personality in the 1990s led to wider acceptance of personality testing in hiring amongst I-O psychologists, albeit not without controversy. (See Section~\ref{sec:audit:context} for more on the Big Five.)

The use of a traditional personality test in personnel selection relies on the following assumptions:
\begin{itemize}
    \item the personality traits being measured are meaningful constructs;
    \item the test is a valid measurement instrument: it measures the traits it purports to measure;
    \item the test is a valid hiring instrument: its results are predictive of employee performance.
\end{itemize}

\paragraph{Validity and reliability of psychometric instruments.} Within the field of psychometrics, instruments are considered useful only if they are both reliable and valid~\cite{cardinet_symmetry_1976,carmines_reliability_1979}. \emph{Reliability} refers to the consistency of an instrument’s measurements, and \emph{validity} is the extent to which the instrument measures what it purports to measure~\cite{mueller_reliability_2018}. Reliability is a necessary (although not a sufficient) condition for validity~\cite{nunnally_psychometric_1994}. Thus, when considering psychometric instruments, the question of reliability is central to the question of validity.

Reliability can be measured across time (\emph{test-retest reliability}), across equivalent forms of a test (\emph{parallel forms reliability}), across testing environment (\emph{cross-situational consistency}), \etc (Mueller and Knapp 2018\nocite{mueller_reliability_2018}). Each of the dimensions across which measurements are compared is referred to as a ``facet,'' such that we can talk about reliability with respect to some facet (\eg time) that varies between measurements, while other facets (\eg test location) are held constant~\cite{cardinet_symmetry_1976}. Under Classical Test Theory (CTT), measurements can be decomposed into a true score and a measurement error~\cite{schmidt_beyond_2003}. The true score is the value of the underlying construct of interest (\eg extraversion). Measurement error can be further broken down across various experiment facets~\cite{schmidt_beyond_2003}. 

Reliability is usually measured and evaluated with correlations. Although 0.80 is often cited as an acceptable threshold of reliability, Nunnally and Bernstein~\cite{nunnally_psychometric_1994} differentiate between standards used to compare groups (for which 0.80 is an appropriate reliability), and those used to make decisions about individuals. For the latter type of test, they advise that 0.90 should be the ``bare minimum,'' and that 0.95 should be the ``desirable standard.''

\paragraph{Algorithmic personality tests,} on which we focus in this paper, constitute a category of psychometric instruments, and are thus relying on the same assumptions---about test validity as a measurement instrument and as a hiring instrument---as do their traditional counterparts.  Guzzo \etal~\cite{guzzo_big_2015} caution that reliability and validity are ``often overlooked yet critically important'' in big-data applications of I-O psychology.  In our work, we aim to fill this gap by interrogating the reliability of algorithmic personality predictors.  Because the objects of our study are algorithmic systems that are used by employers in their talent acquisition pipelines, our work falls within the domain of hiring algorithm audits, discussed next. 

\subsection{Auditing of Hiring Algorithms}
\label{sec:background:audits}

\paragraph{Background on algorithm auditing.} The algorithm audit is a crucial mechanism for ensuring that AI-supported decisions are \emph{fair}, \emph{safe}, \emph{ethical}, and correct. Increasing demand for such audits has led to the emergence of a new industry, termed Auditing and Assurance of Algorithms by Koshiyama \etal~\cite{koshiyama_towards_2021}. 

Scholarly work on algorithm auditing acknowledges that auditing frameworks are inconsistent in terms of scope, methodology, and evaluation metrics~\cite{bandy_problematic_2021,brown_algorithm_2021,koshiyama_towards_2021,raji_closing_2020}. In this landscape that offers many frameworks, yet minimal technical guidance, auditors are left to define their own scope. As argued by several authors, stakeholder interests should be central to the task of scoping~\cite{brown_algorithm_2021,fjeld_principled_2020,metcalf_algorithmic_2021,ORCAA,raghavan_mitigating_2020,raji_closing_2020,razavi_future_2021,sloane_silicon_2021,vecchione_algorithmic_2021}. Sloane \etal ~\cite{sloane_silicon_2021} argue that audits ought to be specific to the domain and to the tool under study.

\revv{In the United States,} much of the audit literature surrounding predictive hiring technology is significantly concerned with legal liability as laid out in the Uniform Guidelines on Employee Selection Procedures (UGESP)~\cite{kim_data-driven_2017,raghavan_mitigating_2020,wilson_building_2021}. These guidelines, adopted by the Equal Employment Opportunity Commission in 1978~\cite{UGESP}, revolve around a form of discrimination called disparate impact, wherein a practice adversely affects a protected group of people at higher rates than privileged groups. As a result, audits of AI hiring systems are often specifically concerned with adverse impact~\cite{chen_investigating_2018,ORCAA,wilson_building_2021}. It is often noted that avoiding liability is not actually sufficient to ensure an ethical system; that is, a lack of adverse impact should be a baseline rather than the goal~\cite{barocas_big_2016,ORCAA,raghavan_mitigating_2020,wilson_building_2021}.
 
\revv{The main contribution of our work is a socio-technical audit methodology developed to measure the stability of personality prediction systems used in the hiring domain, and an open-source library that generalizes the technical components of this framework for use more broadly in stability auditing. We further instantiate this framework on two real-world personality prediction systems.}  As we will discuss in Section~\ref{sec:methodology}, we build on Sloane \etal~\cite{sloane_silicon_2021} to interrogate the assumptions encoded by these systems. 

\paragraph{Treatment of reliability in algorithm audits.} The audit literature is inconsistent in whether reliability is included as \rev{a} concern and, if it is, how it is defined and treated.  
Specifically, several impactful lines of work do not consider reliability~\cite{hagendorff_ethics_2020,langenkamp_hiring_2020,metcalf_algorithmic_2021,sandvig_auditing_2014,suhr_does_2021,venkatadri_privacy_2018,wilson_building_2021}. Of the works that do take reliability under consideration, some refer to this concept as ``stability''~\cite{brown_algorithm_2021,koshiyama_towards_2021,robertson_auditing_2018,sloane_silicon_2021,supreme}, some refer to it as ``reliability''~\cite{fjeld_principled_2020,mokander_ethics-based_2021,raji_closing_2020,shneiderman_bridging_2020,supreme}, and some refer to it as ``robustness''~\cite{chen_investigating_2018,fjeld_principled_2020,mokander_ethics-based_2021,oala_ml4h_2020,ORCAA}. Bandy~\cite{bandy_problematic_2021} forgoes specific terminology and simply refers to changes to input and output.   This difference in treatment is more than terminological: stability relates to local numerical analyses, whereas robustness tends to refer to broad, system-wide imperviousness to adversarial attack, and reliability connotes consistency and trustworthiness.

This inconsistency is part of a larger problem within sensitivity analysis --- the formal study of how system inputs are related to system outputs. Razavi \etal~\cite{razavi_future_2021} observe that sensitivity analysis is not a unified discipline, but is instead spread across many fields, journals and conferences, and notes that lack of common terminology remains a barrier to unification. In our work, we use the term \emph{stability} to refer to a property of an algorithm whereby small changes in the input lead to small changes in the output. We adopt a psychometric definition of \emph{reliability}, which we use to guide the way in which we measure stability. By considering algorithms within their sociotechnical context, we can also translate between numerical stability and broader \emph{robustness}.

Although reliability has not been centered in algorithm audits, the importance of model stability has long been established~\cite{turney_technical_1995}. The 2020 manifesto on responsible modeling by Saltelli \etal~\cite{saltelli_five_2020} underscores the importance of sensitivity analysis, and both the European Commission~\cite{EC} and the European Science Academies~\cite{SAPEA} have called for sensitivity auditing in the policy domain. As detailed by Razavi \etal~\cite{razavi_future_2021}, sensitivity audits have also been applied in the domains of education~\cite{araujo_pisa_2017}, food security~\cite{saltelli_problematic_2017}, public health~\cite{Piano}, and sustainability~\cite{galli_questioning_2016}. We argue that algorithm auditors should consider stability among the critical metrics they select from, as suggested by Brown \etal~ \cite{brown_algorithm_2021}.

Our work is synergistic with two recent lines of work that contribute substantive quantitative methodologies for auditing algorithm stability. In the first, Xue \etal~\cite{xue_auditing_2020} introduce a suite of tools to study individual fairness in black-box models. In the second, Sharma \etal~\cite{sharma_certifai_2020} offer a unified counterfactual framework to measure bias and robustness. Sharma \etal's methodology relies on access to the features being used by the model, whereas the methods proposed by Xue \etal and by our work only require query access to black-box models. The key distinction between Xue \etal and our work is that Xue \etal build on notions of individual fairness that can be encoded by Wasserstein distance, while we approach stability through a sociotechnical lens, borrowing metrics that are familiar to I-O psychologists.

\paragraph{Audit scope.} 
\rev{A} number of \rev{recent} algorithm audits \rev{focus} on tools used at various stages in hiring pipelines. Wilson \etal~\cite{wilson_building_2021} and O'Neil Risk Consulting and Algorithmic Auditing (ORCAA) ~\cite{ORCAA} each \rev{focus} on tools for pre-employment assessment (\ie candidate screening). Raghavan \etal~ \cite{raghavan_mitigating_2020} \rev{evaluate} the public claims about bias made by the vendors of 18 such tools. Chen \etal~\cite{chen_investigating_2018} \rev{audit} three resume search engines, Hann{\'a}k \etal~\cite{hannak_bias_2017} \rev{audit} two online freelance marketplaces, and De-Arteaga \etal~\cite{de-arteaga_bias_2019} \rev{builds} and \rev{evaluates} several classifiers that predict occupation from online bios. All of these studies focus primarily on bias and discrimination. It \rev{is also} common to frame these audits around the promises that companies make in their public statements~\cite{ORCAA,raghavan_mitigating_2020,wilson_building_2021}. By contrast, in our work we focus on auditing stability, which is a necessary condition for the validity of an algorithmic hiring tool.

Access level is a critical factor in determining audit scope. Audits can be internal (where auditors are employed by the company being audited), cooperative (a collaboration between internal and external stakeholders), or external (where auditors are fully independent and do not work directly with vendors). Sloane \etal~\cite{sloane_silicon_2021} explain that the credibility of internal audits must be questioned, because it is advantageous to the company if they perform well in the audit. Ajunwa~\cite{ajunwa_auditing_2021} argues for both internal and external auditing imperatives, with the latter ideally performed by a new certifying authority.  Brown \etal~\cite{brown_algorithm_2021} offer a flexible framework for external audits that centers on stakeholder interests.  Bogen and Rieke~\cite{bogen_help_2018} stress the importance of independent algorithm evaluations and place the burden on vendors and employers to be ``dramatically'' more transparent to allow for rigorous external audits. Absent that transparency, however, external audits must be designed around what information is publicly available.  In this work we develop an external auditing methodology.  

\section{Methodology}
\label{sec:methodology}

\subsection{Socio-technical methodology}

We now present a socio-technical framework to assess the stability of algorithmic personality tests in hiring, inspired by the  auditing framework of Brown \etal~\cite{brown_algorithm_2021}. 

\begin{enumerate}
\item  \textbf{Define the socio-technical context} in which the system operates, and detail the system's inputs and outputs.

\item \textbf{Identify assumptions} made by the vendors regarding stability of the system.

\item \textbf{Identify key facets} of measurement across which the system assumes its outputs to be stable, based on validity assumptions.

\item \textbf{Collect data} that is representative of the tool's intended context of use.

\item \textbf{Generate treatments} by perturbing the input (control) across the features that correspond to each facet of measurement, while keeping all other features fixed to the extent possible.

\item \textbf{Identify stability metrics and acceptance/rejection criteria} that suitably capture the statistical relationship between the control and treatments.

\item \textbf{Query the external system of interest} to collect scores for the control and treated inputs.

\item \textbf{Quantify the \emph{instability} across each facet} based on the selected statistical criteria. 

\end{enumerate}

\label{sec:framework}

\begin{figure}
\centering 
\includegraphics[width=0.9\textwidth]{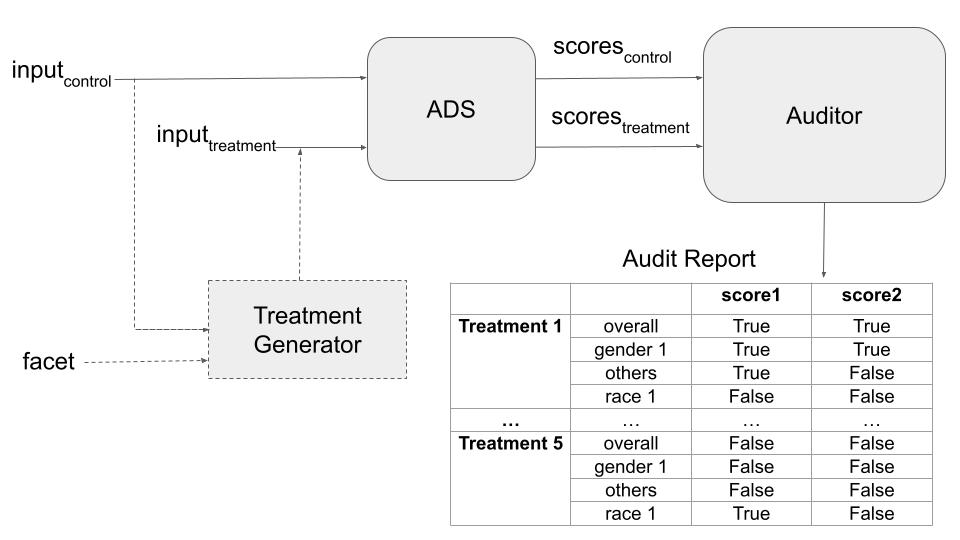}
\caption{\revv{Overview of the technical framework, implemented by our open-source library.}}
\label{fig:technical}
\end{figure}

\subsection{Technical framework and open-source library}

As part of this work, we developed an open-source software library \footnote{https://github.com/DataResponsibly/hiring-stability-audit} that implements the technical components of the audit, and can be used for the stability audit of any automated decision system (ADS), given suitable input data, treatment generation techniques, and choice of stability metrics. The technical framework is shown in  Figure~\ref{fig:technical} and consists of three modules: ADS, Treatment Generator, and Auditor, described next.

\subsubsection{ADS}

The ADS class is an abstraction of the algorithmic system being studied.  It has a generic \emph{score function} that takes inputs and returns scores. We include the ADS class in the audit framework to make explicit the nature of access that the user has to the system being studied, and it is intentionally designed to be generic to allow the user to model different auditing conditions. For example, for external audits that treat the system as a black box, the score function of the ADS object will be a simple look-up of the outputs that were produced by the system.  As another example, external audits that do not have direct access to the system may instead fit a model on the collected data.  Then, the score function will be executed over that fitted model.  The framework can similarly be applied for internal audits, where the user has access to the model and invokes the score function directly. 

Users can also implement custom score functions to generate \emph{treatment baselines}.  We would expect any algorithmic system that is used in the real world to be at least as good as a random guesser, and so a score function that appropriately implements random guessing can be used as a baseline for model stability, where the variation between control and treatment scores for any of the treatments should not exceed the variation observed between the control and the randomly generated outputs. 

\subsubsection{Treatment Generator}
In this work, we focus on an important desideratum of algorithmic systems --– stability. We measure stability based on how robust system outputs are to different \emph{treatments} performed on the input. The treatment generator is a technical instantiation of the mechanism that generates treatments, based on a particular \emph{facet of measurement} of stability. In the audits implemented in this study, described in Section~\ref{sec:audit}, the treatments are hand-designed based on domain expertise with personality scores and their use in hiring.  However, we envision that future technical frameworks can at least partially automate the creation of treatments, for example, by sampling values for a particular feature from an appropriate distribution, or by automatically perturbing values in text features. 

\subsubsection{Auditor}
The Auditor is the main class of this technical framework: it instantiates a generic Auditor that can be used to analyze the stability of a variety of algorithmic systems.   The generic Auditor class allows the user to specify the following information:
\begin{enumerate}
    \item \textbf{Score names.} The framework is flexible to test the stability of multiple scores produced by the same system, and takes in a list of score names from the user. For example, \crystal  produces four scores corresponding to the personality traits of 'Dominance',  'Influence', 'Steadiness', and 'Conscientiousness’.
    \item \textbf{Control score.} This is the score corresponding to the unmodified/unperturbed input.
    \item \textbf{Treatment scores.} The framework is flexible to simultaneously analyze the stability of the ADS for several different treatments, and accepts a dictionary of treatment names and the corresponding treatment scores from the user. For example, in the audit of Crystal, treatments include ``modifying input type'', ``modifying source context'', ``embedding LinkedIn URL'', etc. 
    \item \textbf{Demographic information.} The framework can break down audit results by demographic group. To invoke this functionality, the user can pass identifiers that link demographic information to each input and score, and specify \emph{groups of interest} within the population being evaluated/scored by the ADS. All the subsequent analysis is then performed for both the overall population and for the specified groups. 
\end{enumerate}

\noindent The Auditor class currently supports the following functionality:

\paragraph*{\textbf{Statistical hypothesis testing.}}~~From a socio-technical standpoint, this audit framework identifies facets across which the system assumes its output to be stable, and then tests the validity of those claims. The socio-technical heavy-lifting is in identifying these facets, designing treatments that vary along this facet, and identifying measures that capture the variation between control and treatments as a measure of instability. The Auditor class of the technical framework automates the subsequent hypothesis testing by instantiating a generic \emph{compute$\_$statistic()} method. This method supports several popular correlation tests (such as Spearman, Pearsons and Kendall-Tau), parametric tests (such as student-t, paired student-t and ANOVA) and non-parametric tests (such as Mann-Whitney, Wilcoxon and Kruskal-Wallis). Users can also choose to plug in their own custom functions to compute statistics that the framework does not currently support.

Users can thereby test their hypotheses about the stability of the ADS with respect to different treatments, and for demographic groups of interest, by using the Auditor class with the relevant statistical measure.

\paragraph*{\textbf{Measuring total variation.}}~~Our audit framework supports functionality to compute and visualize the total variation between the control and treatments, with a large amount of variation indicating greater instability. The generic \emph{compute$\_$total$\_$variation} method of the Auditor class implements this. For the purposes of the \crystal and \humantic audits, we chose to measure  total variation as the L1 distance between the control and each treatment, but the framework is flexible to accommodate different measures of total variation. 

The Auditor class also has a \emph{visualize$\_$total$\_$variation} method that produces box plots of the total variation for each treatment, broken down by demographic groups of interest. Extensions of this framework could include additional measures and visualization techniques to analyze the total variation.  

\paragraph*{\textbf{Visualization.}}~~The Auditor class implements the \emph{visualize$\_$scores} method that produces scatter plots of control vs. treatment scores, compared with the ideal $Y=X$ line (no variation between control and treatment scores), and the \emph{visualize$\_$total$\_$variation} method discussed above.  

\section{Instantiation of the framework for personality testing in hiring}
\label{sec:audit}

\revv{We now instantiate our socio-technical framework through external audits of \humantic and \crystal. Jupyter notebooks demonstrating the use of our open-source library to conduct these audits are available on our \footnote{https://github.com/DataResponsibly/hiring-stability-audit}{Github}. It also contains a third audit on a synthetically-generated dataset to demonstrate a broader, more general application of the technical framework.}

\subsection{Socio-technical context}
\label{sec:audit:context}
Employers purchase candidate-screening tools from \crystal and \humantic and use them to build personality profiles of potential employees. Both systems offer functionality for ranking candidates based on their personality profiles. \crystal assigns a ``job fit'' score to candidates, which is measured based on a comparison to either a ``benchmark candidate'' with a user-specified ideal personality profile, or to a job description that is analyzed to ``detect the most important personality traits.'' Similarly, \humantic assigns a ``match score'' to candidates by comparing them to an ``ideal candidate,'' specified with a LinkedIn URL or an ideal personality score vector. 

The hiring processes supported by these systems are not fully automated. Human decision-makers must choose whether and how to define an ideal candidate, at what stage of hiring to use the tool, and how to incorporate tool outputs into hiring decisions. For example, an HR professional may decide to use an existing employee to define an ideal candidate, then run all resumes they receive through the tool, and finally offer interviews to all candidates with match scores above 90\%. A different HR department may use the system to filter resumes before human review, choosing to rank candidates based on predicted ``Steadiness'' scores, and then discard all but the top 25 candidates. As these \rev{examples} illustrate, the human-in-the-loop implementation details are crucial to actual outcomes.

\paragraph{\textbf{Inputs and outputs.}}
Both systems output candidate DiSC scores: vectors of 4 numeric values, each corresponding to a personality trait. \humantic produces a score for each trait on a scale from 0 to 10, while \crystal represents each trait as a percent of the whole, giving each a score from 0 to 100 such that all four traits sum to 100\%. In addition to DiSC, \humantic also outputs scores for The Big Five model of personality. 

DiSC is a behavioral psychology test that assesses the extent to which a person exhibits four personality traits: Dominance (D), Influence (I), Steadiness (S), and Conscientiousness (C).\footnote{\url{https://www.discprofile.com/what-is-disc/how-disc-works}} Although official DiSC documentation states that C represents ``Conscientiousness,'' \humantic states that C in DiSC stands for ``Calculativeness.''\footnote{\humantic separately produces predictions on ``Conscientiousness'' within the Big Five model of personality. We posit that \humantic may have made the choice to rename the DiSC ``Conscientiousness'' trait to ``Calculativeness'' in order to avoid conflation with the Big Five trait by the same name. } 
Notably, although both \humantic and \crystal market DiSC as a rigorous psychology-based analysis methodology, scholarly work on DiSC in I-O psychology has been limited, especially with regard to its validity and reliability for hiring. In fact, the DiSC website explicitly states that DiSC scores are ``not recommended for pre-employment screening.''\footnote{\url{https://www.discprofile.com/everything-disc/hiring}}

The Big Five model is far better studied than DiSC, and its use in personnel selection is considered acceptable by some I-O psychologists~\cite{goodstein_applications_1999,hurtz_personality_2000}). Still, the use of the Big Five in hiring is not without criticism. For example, Morgeson \etal~\cite{morgeson_reconsidering_2007} argue that ``the validity of personality measures as predictors of job performance is often disappointingly low.''  The Big Five model contains five traits: Openness (O), Conscientiousness (C), Extraversion (E), Agreeableness (A), and Neuroticism (N). \humantic replaces Neuroticism with the more palatable ``Emotional Stability'', which, they explain, is ``the same as Neuroticism rated on a reverse scale.''\footnote{\url{https://app.humantic.ai/\#/candidates}} 

\paragraph{\textbf{System design and validation.}} \humantic and \crystal state that they use machine learning to extract personality profiles of job candidates based on the text of their resumes and LinkedIn profiles.  However, public information about model design and validation is limited. \humantic states that ``all profile attributes are determined deductively and predictively from a multitude of activity patterns, metadata or other linguistic data inputs.''\footnote{\url{https://api.humantic.ai/}} \crystal explains that their personality profiles are ``predicted through machine learning and use text sample analysis and attribute analysis\rev{.''\footnote{\url{https://www.crystalknows.com/blog/crystal-accuracy}}} Neither company makes its training data publicly available or discusses the data collection and selection methodology they used.  For this reason, an external audit cannot assess whether the training data is representative of the populations on which the systems are deployed. 

Information about validation is limited as well. \humantic reports that their outputs ``have an accuracy between 80-100\%''\footnote{\rev{\url{https://api.humantic.ai/}}} \crystal advertises that ``based on comparisons to verified profiles and our user’s direct accuracy validation through ratings and endorsements, Crystal has an 80\% accuracy rating for \rev{P}redicted \rev{[sic]} profiles.''\footnote{\url{https://www.crystalknows.com/blog/crystal-accuracy}} No additional information is given about the validation methodology, the specific accuracy metrics, or results.  Finally, update schedules for the models used by the systems are not disclosed.

\subsection{System assumptions}
\label{sec:audit:assumptions}
In accordance with Sloane \etal~\cite{sloane_silicon_2021}, our methodology is centered around testing the underlying assumptions made by algorithmic systems within their specific socio-technical context. Because algorithmic personality tests constitute a category of psychometric instrument, they are subject to the assumptions made by the traditional  instruments, as laid out in Section~\ref{sec:psychometric_theory}. The validity of these systems is subject to the following additional assumptions: \footnote{\rev{Note that this list of assumptions is not exhaustive.}}  
\begin{itemize}  
    \item [{\bf A1:}] The output of an algorithmic personality test is stable across input types (such as PDF or \rev{Docx}) and other job-irrelevant variations in the input.  This assumption corresponds to parallel forms reliability from psychometric testing (see  Section~\ref{sec:psychometric_theory}).
    \item [{\bf A2:}]  The output of an algorithmic personality test is stable across input sources (such as resume or LinkedIn) that are treated as interchangeable by the vendor.  This assumption corresponds to  cross-situational consistency (see  Section~\ref{sec:psychometric_theory}).
    \item [{\bf A3:}] The output of an algorithmic personality test on the same input is stable over time. This assumption corresponds to  test-retest reliability (see  Section~\ref{sec:psychometric_theory}).
\end{itemize}

Importantly, all these assumptions are testable via an external audit. Thus, these are the assumptions on which we focus our analysis, and with respect to which we quantify stability as a necessary condition for validity.

\subsection{Key facets of measurement}
\label{sec:audit:facets}
We identify the following key facets across which \humantic and \crystal operationalize reliability, as discussed in Section~\ref{sec:methodology}:

   \paragraph*{\textbf{Resume file format.}}~Absent specific formatting instructions, the file format of an applicant's resume (\eg PDF or text), should have no impact on their personality score.  Per assumption {\bf A1}, stability estimates across this facet quantify parallel forms reliability.

     \paragraph*{\textbf{Source context.}}~Both systems use implicit signals within certain contexts (\ie resumes, LinkedIn profiles, and tweets) to assign personality scores to job seekers.  Further, both systems allow direct comparisons of personality scores derived from multiple source contexts, for example by ranking candidates on their ``match score,'' which is computed from resumes for some job seekers and from LinkedIn profiles for other job seekers. Per assumption {\bf A2}, stability estimates across this facet quantify cross-situational consistency.
    
     \paragraph*{\textbf{Inclusion of LinkedIn URL in a resume.}}~The decision to embed a LinkedIn URL into one's resume should have no impact on the personality score computed from that resume.  This is because output is expected to be stable across input sources per assumption {\bf A2}, and across job-irrelevant input variations per {\bf A1}.  

     \paragraph*{\textbf{Algorithm-time}}~(time when input is scored). Both systems generate personality scores for the same input at different points in time, and they compare and rank job seekers based on their scores made at different times. For example, consider an extended hiring process that takes place over the course of months, with new candidates being screened at different times. In this situation, \humantic and \crystal would both encourage users to compare output generated months apart.  Based on assumption {\bf A3} (test-retest reliability), we expect the personality score computed on {\em the same input} to be the same, irrespective of when it is computed.
    
     \paragraph*{\textbf{Participant-time}}~(time when input is produced). An employer may keep candidate resumes on file to consider them for future positions. An HR specialist might be tempted to generate scores from resumes they have on file, and compare them to scores of new candidates. Neither \humantic nor \crystal offer any guidance to users regarding the time period during which results remain valid, thus encouraging users to generalize across participant-time. Based on {\bf A3} (test-retest reliability), we expect the personality score computed based on time-varying input from \emph{the same individual} to be the same, irrespective of when the input is generated.
   
\subsection{\revv{Data collection}}
\label{sec:inputs_and_experiments}

\paragraph{Primary data collection.}  We conducted an IRB-approved human subjects research study at New York University to seed the input corpus for the audit. For this, we recruited current graduate students at New York University’s Center for Data Science ($N=33$), Tandon School of Engineering ($N=51$), and Courant Institute of Mathematical Sciences ($N=10$). We further required that participants not be currently located in the European Union or the United Arab Emirates. Participants were asked to complete a survey to upload their resume, provide a link to their public LinkedIn URL, their public Twitter handle, and their demographic information. All survey questions were optional. 

In total, 94 participants qualified for the study, of whom 92 submitted LinkedIn URLs, 89 submitted resumes (in PDF, Microsoft Docx, or .txt format), and 32 submitted public Twitter handles. Participants were given access to their personality profiles computed by \crystal and \humantic in exchange for their participation in the study. 88\% of participants were pursuing a \rev{Master's} Degree and 12\% were pursuing a PhD degree. 60\% of participants identified as male, 38\% as female, and 2\% as non-binary. Their ages ranged from 21-40 with a mean of 26.13. 60\% of our sample identified as Asian, 25\% as White, 5\% as Hispanic \rev{or} Latino, 3\% as Black or African American, and 4\% identified as two or more races. 1\% declined to identify their race. 37\% of participants were born in India, 30\% were born in the US, 13\% were born in China, and 20\% were born elsewhere. 64\% reported that English was their primary language. (See Appendix~\ref{app:details:demographics}.)

\paragraph{Persistent linkage of email addresses to LinkedIn profiles, and the need for de-identification.} During the initial processing of participant information in \humantic, we observed that the personality profile produced from LinkedIn is often identical to the one produced from a resume containing an embedded LinkedIn URL. We hypothesized that for such URL-embedded resumes, \humantic was disregarding any information on the resume itself and  pulling information from LinkedIn to generate a personality score. We further hypothesized that the system may create persistent linkages between email addresses and LinkedIn profiles. 

\begin{table*}[t]
\small 
\caption{Resume versions used as input.}
\label{resume_versions}
\begin{tabular}{p{0.13\textwidth} | p{0.1\textwidth} | p{0.65\textwidth}}
\hline\noalign{\smallskip}
Version & File Format & Pre-Processing  \\
\noalign{\smallskip}\hline\noalign{\smallskip}
Original & Various & None \\
De-Identified & PDF & Remove identifiers (name, phone, email, social media links, usernames). Save as PDF. \\
Raw Text & Raw Text & Copy text. \\
PDF & PDF & Save as PDF (if original in other format). \\
DOCX & DOCX & Remove identifiers (name, phone, email, social media links, usernames). Save as DOCX. \\
URL-Embedded & PDF & Remove identifiers (name, phone, email, social media accounts, LinkedIn URL).  Insert hyperlinked LinkedIn URL into beginning of document. Save as PDF. \\
\noalign{\smallskip}\hline
\end{tabular}
\end{table*}
To investigate this trend, resumes containing a LinkedIn URL and an email address were passed to \humantic. Next, we created and submitted fake PDF ``resumes,'' which were blank except for the email addresses that had been passed along with LinkedIn URLs, and compared the \humantic output produced by these two treatments. (Note: Due to privacy concerns, all linkage experiments used researchers' own accounts and either their own or synthetic email addresses.)  It was revealed that, when \humantic encounters a document that contains both a LinkedIn URL and an email address, it persistently associates the two such that the system produces the same personality score whenever it encounters that email address in the future. 
Because \humantic uses the embedded URLs to import information directly from LinkedIn, the predicted profiles in our linkage experiments displayed names, photos, and employment information present on LinkedIn, but not on the resumes.   

These findings further substantiate that \humantic operationalizes assumption \textbf{A2} of cross-situational consistency (see Section~\ref{sec:methodology}).

These findings necessitated the use of de-identified resumes in all future \humantic experiments. De-identification allows comparison of the algorithm's predictions on resumes, without the obfuscating effect of information being pulled from LinkedIn. It also prevents participants' emails \rev{from} being linked to synthetically altered versions of their resumes. See Table~\ref{resume_versions} for de-identification details. Note that de-identification was not necessary in \crystal, as no such linkage was observed there. Further findings from our linkage explorations are detailed in Section~\ref{sec:linkage_results}.

\subsection{\revv{Treatment generation}}
To assess stability with respect to a facet of measurement, we need to perturb the input across the features that correspond to each facet, while keeping all other features fixed to the extent possible.  As a result, we generate a pair of datasets, which we call \emph{treatments}, for each facet.  To isolate facet effects as cleanly as possible, we prepared several resume versions, described in Table~\ref{resume_versions}.  Details of each set of score-generating model calls that use these resume versions, or social media links, are presented in Appendix~\ref{app:details:calls}.  We will explain how these versions are used as treatments in the stability experiments in Section~\ref{sec:results}.

\subsection{\revv{Stability measures}}
In the context of personality prediction, we identify the following measures of stability:

    \paragraph*{\textbf{Rank-order stability.}} As explained in Section~\ref{sec:psychometric_theory}, the reliability of psychometric instruments is measured with correlations. Thus, we select correlation as the statistical measure of rank-order stability. Morrow and Jackson~\cite{morrow_how_1993} make a convincing argument against providing significance levels for reliability correlations. Instead, we use the ``bare minimum'' of 0.90 and the ``desirable standard'' of 0.95, as proposed by Nunnally and Bernstein~\cite{nunnally_psychometric_1994} as the accept/reject threshold on correlations.
    
    \paragraph*{\textbf{Locational stability}} If a system allows users to compare output across a key facet, then we should also assess locational stability across that facet, \ie whether one facet treatment generally yields higher overall scores. We select the Wilcoxon signed\rev{-}rank test, a non-parametric alternative which tests whether the median of the paired differences is significantly different than zero, as the statistical measure of locational stability. We select a suitable significance threshold after correcting for multiple hypothesis testing: 
    
    \begin{itemize}
        \item \textbf{Bonferroni correction} controls the family-wise error rate. It is guaranteed to falsely reject the null hypothesis no more often than the nominal significance level, however, it can be overly conservative, especially when sample sizes are low (\ie it can falsely accept the null hypothesis more often than the nominal significance level implies)~\cite{bonferroni}.
        $$\alpha_{\text{Bonferroni}} = \frac{\alpha_{\text{nominal}}}{\text{\# tests performed}}$$
        \item \textbf{Benjamini-Hochberg correction} is a less conservative approach that controls the false discovery rate. The procedure ranks obtained p-values in ascending order and uses these ranks to derive corrected thresholds, which range between $\alpha_{\text{Bonferroni}}$ and $\alpha_{\text{nominal}}$~\cite{benjamini_hochberg}.
        $$\alpha_{\text{Benjamini-Hochberg}} = \frac{\text{p-value rank}}{\text{\# tests performed}}\alpha_{\text{nominal}}$$
    \end{itemize}
  
    \paragraph*{\textbf{Total change}} We also identify total change as a relevant measure of instability, and use the L1 distance to measure it.

\revv{Note that these are three different ways to quantify stability, and that a system may, for example, be found to have sufficient rank-order stability but to lack locational stability, and vice versa.}

\subsection{Generating outputs}
To conduct this audit, we purchased nine months of \humantic basic organizational membership at a total cost of \$2,250, and a combination of monthly and annual \crystal memberships at a total cost of \$753.82. We carried out our experiments over the period of November 23, 2020 through September 16, 2021.

One week into our evaluation, representatives from \humantic ascertained that we were using their tool to conduct an audit, and reached out to inform us that they would like to collaborate in the effort. In light of this development, we weighed the advantages and disadvantages of engaging with \humantic and decided to continue with a neutral external audit, to minimize the potential for conflicts of interest and maximize our ability to critically analyze the system for stability. The cost of that decision is that we had to forgo potential access to the underlying data, modeling decisions, features, and model parameters that a collaboration with \humantic may have afforded~\cite{koshiyama_towards_2021,sloane_silicon_2021}. While we do not have any reason to believe that the discovery of our audit caused \humantic to change their models or operation, we cannot rule out this possibility.

\subsection{Computing stability measures}
For the audits of \crystal and \humantic we compute the following statistical measures using our technical framework:

\paragraph*{\textbf{Rank Order Stability.}}~We compute Spearman's correlation (a measure of rank order stability) as follows:


\begin{Python}[frame=none,captionpos=b,texcl=true]
# Instantiate the Auditor class
stability_audit = Auditor(control_scores, treatment_scores)

# Call the generic compute statistic method with parameter test set to Spearman
corr_ = stability_audit.compute_statistic(test=spearman)["correlations"]

# Threshold correlations on desired cut-offs
corr_threshold = 0.9
corr_ > corr_threshold
\end{Python}

\revv{\paragraph*{\textbf{Locational Stability.}}~We perform Wilcoxon's signed rank test (as the measure of locational stability) as follows:}

\begin{Python}[frame=none,captionpos=b,texcl=true]
# This time calling compute statistic method with parameter test set to Wilcoxon

pvals = stability_audit.compute_statistic(test=wilcoxon)["p_values"]

# Using an alpha of 0.05
alpha_threshold = 0.05

# Correct for multiple hypothesis testing using Benjamini-Hochberg correction
corrected = stability_audit.multiple_hypothesis_correction(
                             pvals, alpha = alpha_threshold, method='fdr_bh')

# Threshold pvalues on desired cut-off
corrected > alpha_threshold

\end{Python}

\revv{\paragraph*{\textbf{Total Variation.}}~We also compute the L1 distance (as a measure of total variation) as follows:}
\begin{Python}[frame=none,captionpos=b,texcl=true]
# This time calling the compute total variation method, whose default measure is the L1 norm

total_variation = stability_audit.compute_total_variation()

\end{Python}
\color{black}
\section{\revv{Results}}
\label{sec:results}



\begin{table}[b]
\centering
\caption{Summary of stability results for \crystal and \humantic, with respect to facets of measurement from Section~\ref{sec:audit:facets}. ``\cmark'' indicates \revv{both sufficient rank-order stability ($r\geq0.90$) and sufficient locational stability ($p \geq \alpha_{\text{Benjamini-Hochberg}}$) in all traits, ``\xmark'' indicates either insufficient rank-order stability ($r<0.90$) or significant locational instability ($p < \alpha_{\text{Benjamini-Hochberg}}$)} in at least one trait, and ``\qmark'' indicates the facet was not tested in our audit.} 
\label{tab:res_summary}
    \begin{tabular}{l| c | c | l}
    \hline 
    Facet & \crystal & \humantic & Details \\
    \hline
    Resume file format & \xmark  & \cmark & Sec.~\ref{sec:file_format_results} \\
    LinkedIn URL in resume & \qmark  & \xmark & Sec.~\ref{sec:url_embedding_results} \\
    Source context &  \xmark &  \xmark & Sec.~\ref{sec:source_context_results}\\
    \rev{Algorithm-time / immediate} & \cmark & \cmark & Sec.~\ref{sec:reproducibility_and_algorithm_time_results}\\
    \rev{Algorithm-time / 31 days} & \cmark  &  \rev{\xmark} &  Sec.~\ref{sec:reproducibility_and_algorithm_time_results}\\
    Participant-time / LinkedIn  & \xmark  & \xmark & Sec.~\ref{sec:participant_time_results}\\
    Participant-time / Twitter  & N/A  & \cmark & Sec.~\ref{sec:participant_time_results}\\
    \hline
    \end{tabular}
    \label{tab:summary}
\end{table}

Table~\ref{tab:res_summary} summarizes the results of our audit.  We found that \humantic and \crystal predictions both exhibit rank-order instability with respect to source context and participant-time. In addition, \crystal is rank-order unstable with respect to file format, and \humantic is rank-order unstable with respect to URL-embedding in resumes. The systems were sufficiently rank-order stable with respect to all other facets. We did not find any significant locational instability in \crystal. Some traits in \humantic displayed significant locational instability with respect to URL-embedding, source context, and participant-time. Complete experimental results can be found in Appendix~\ref{app:res}.

\revv{\subsection{Persistent linkage and privacy violations in \humantic}}
\label{sec:linkage_results}
\revv{Investigative linkage experiments revealed that when \humantic encounters a document that contains a LinkedIn URL and an email address, the resulting profile will have a 100\% confidence score, and it will contain information found only on LinkedIn (including name, profile picture, and job descriptions and dates). Furthermore, the \humantic model produces the same personality profile whenever it encounters that email address in the future. This linkage persists regardless of how different the new resume is from the one that initially formed the linkage. The email address in question need not be associated with the LinkedIn profile, or even with the candidate. We observed one case in which a participant listed contact information for references, and \humantic created a link between a reference's email and the participant's LinkedIn.} 

\revv{We also found that, once a linkage between an email address and a LinkedIn URL had been made, we were able to alter the personality score produced from a LinkedIn profile by submitting a resume with strong language, namely, containing keywords ``sneaky'' and ``adversarial.'' We therefore conclude that the linkage is used by \humantic in both directions: the content of a LinkedIn profile can affect the personality score computed from a linked resume, and the content of a linked resume can affect personality score computed based on a LinkedIn profile.}

\revv{We did not observe any linkage with participants' Twitter accounts. However, when we used high-profile celebrity Twitter accounts as input, \humantic produced profiles that contained links to several other profiles, including Google+, LinkedIn, Facebook, and Klout. We observed one case in which a high-profile popstar was linked to a software engineer of the same name.}

\begin{figure}
\centering 
\includegraphics[width=0.9\textwidth]{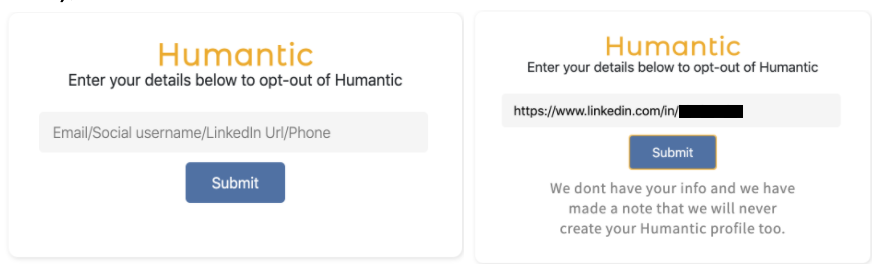}
\caption{Screen shots of the \humantic ``opt out'' feature.}
\label{opt_out}
\end{figure}

\revv{Although \humantic offers an option at the bottom of their website to ``opt out of \humantic'' by entering an email, social network username, LinkedIn URL, or phone number (see Figure~\ref{opt_out}), this feature seems to be inoperable. Various forms of participant information were entered into this field, yet, personality scores associated with this information in the past persisted on the \humantic dashboard, and new results were returned when the information was passed to \humantic in a new account. In cases where LinkedIn profiles were deactivated after profiles were created from them, it was observed that \humantic would still create new profiles from the deactivated LinkedIns, even on different \humantic accounts.} \bigskip

\revv{\subsection{Score distributions}}
\label{sec:results:dist}

\revv{Output scores in \humantic were approximately normally distributed, with the exception of DiSC Calculativeness, which was strongly left-skewed in all runs. }

\begin{figure}
\centering 
\includegraphics[width=0.8\textwidth]{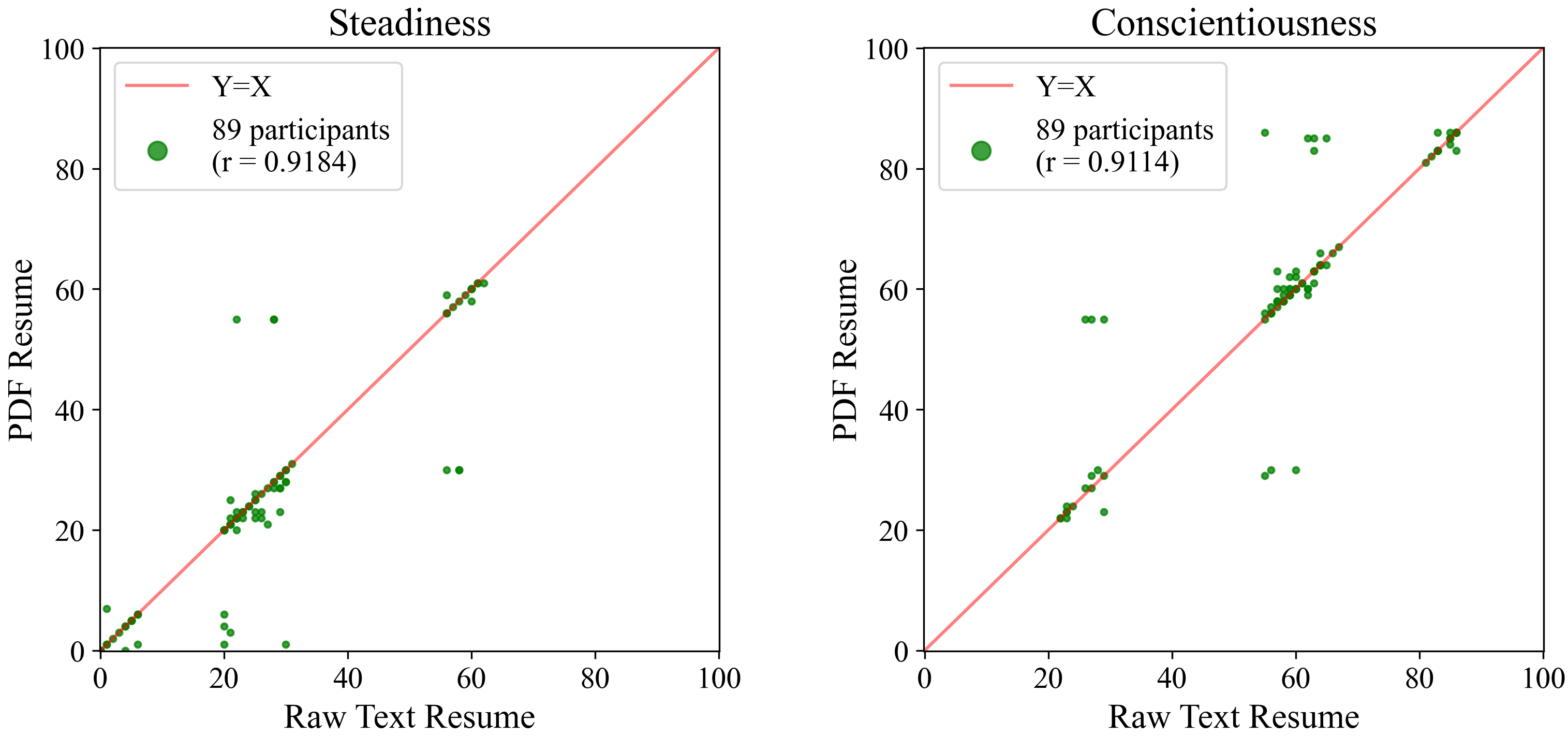}
\caption{Comparison of \crystal output across the resume file format facet.  Note evidence of discontinuous measurement in DiSC Steadiness and Conscientiousness, with some participants' scores moving between clusters with different file formats.}
\label{fig:crystal_discont}
\end{figure}

\revv{We observed discontinuity in \crystal output, which was particularly marked in Steadiness and Conscientiousness, as shown in Figure~\ref{fig:crystal_discont}. For example, no one in our sample had a Steadiness score between 40-50, but many individuals had scores in the 20-30 range, and then again in the 55-65 range.  This may be problematic from the point of view of stability, because a small change in the input may lead to a large change in output across the point of discontinuity, effectively moving between clusters. In fact, we observe this in  Figure~\ref{fig:crystal_discont}, where in two cases, the value of Steadiness jumps from around 30 for raw text resumes to around 60 for PDF resumes.  Having a PDF resume can make you twice as steady, according to \crystal.  Yet, two other examples show the opposite effect: A raw text resume scores about twice as high on Steadiness compared to PDF, for another pair of individuals in our sample.  And so having a PDF resume can also make you half as steady, according to \crystal. There are further examples of this for Conscientiousness, also shown in Figure~\ref{fig:crystal_discont}.}


\revv{We found no evidence of significant locational instability in \crystal. The median for each DiSC trait remained fairly constant across all \crystal runs. The median Dominance score was always 5, the median Influence score was always 10, the median Steadiness score was always 22 or 23, and the median Conscientousness score ranged from 59 to 62.} \bigskip  

\subsection{File format}
\label{sec:file_format_results}

We determine that \humantic is in general sufficiently stable with respect to file format. Rank correlations range from 0.982 (Emotional Stability) to 0.998 (Steadiness). (The two sets of runs are constant with regard to participant-time, and \rev{are} very close to each other in terms of algorithm-time; scores for the de-identified PDF and Docx resumes were generated on the same day, within minutes of each other.)


\revv{\crystal's overall stability across the file format facet fails to meet Nunnally and Bernstein's preferred standard of 0.95 for Steadiness (0.918) and Conscientiousness (0.911), and falls below the minimum limit of 0.90 for Dominance (0.822) and Influence (0.826). In some subgroups, Steadiness and Conscientiousness do fall below 0.90: female ($N=33$) and those whose primary language is English ($N=56$). Although PDF resumes were scored by \crystal four months earlier than raw text resumes, given the perfect reproducibility of \crystal's text  predictions, albeit over a shorter time span, we can assume that algorithm-time is not a factor here.}

There \rev{were} no significant locational stability differences across the file format facet in either \humantic or \crystal. 

\subsection{Inclusion of LinkedIn URL in resume}
\label{sec:url_embedding_results}

\begin{figure}[t!]
\centering
\subfloat[]{
\includegraphics[width = 0.42\columnwidth]{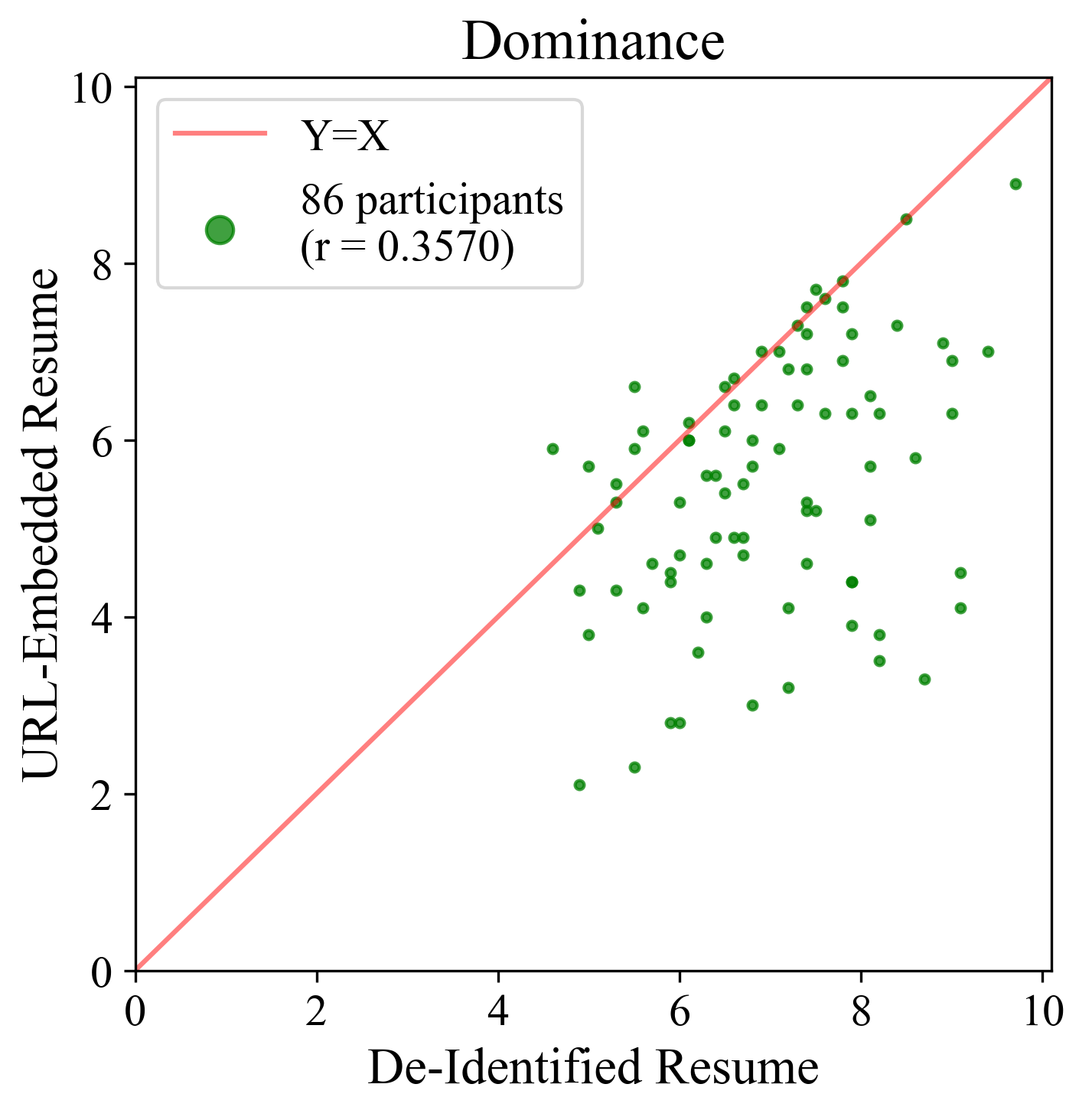}}
\subfloat[]{
\includegraphics[width = 0.45\columnwidth]{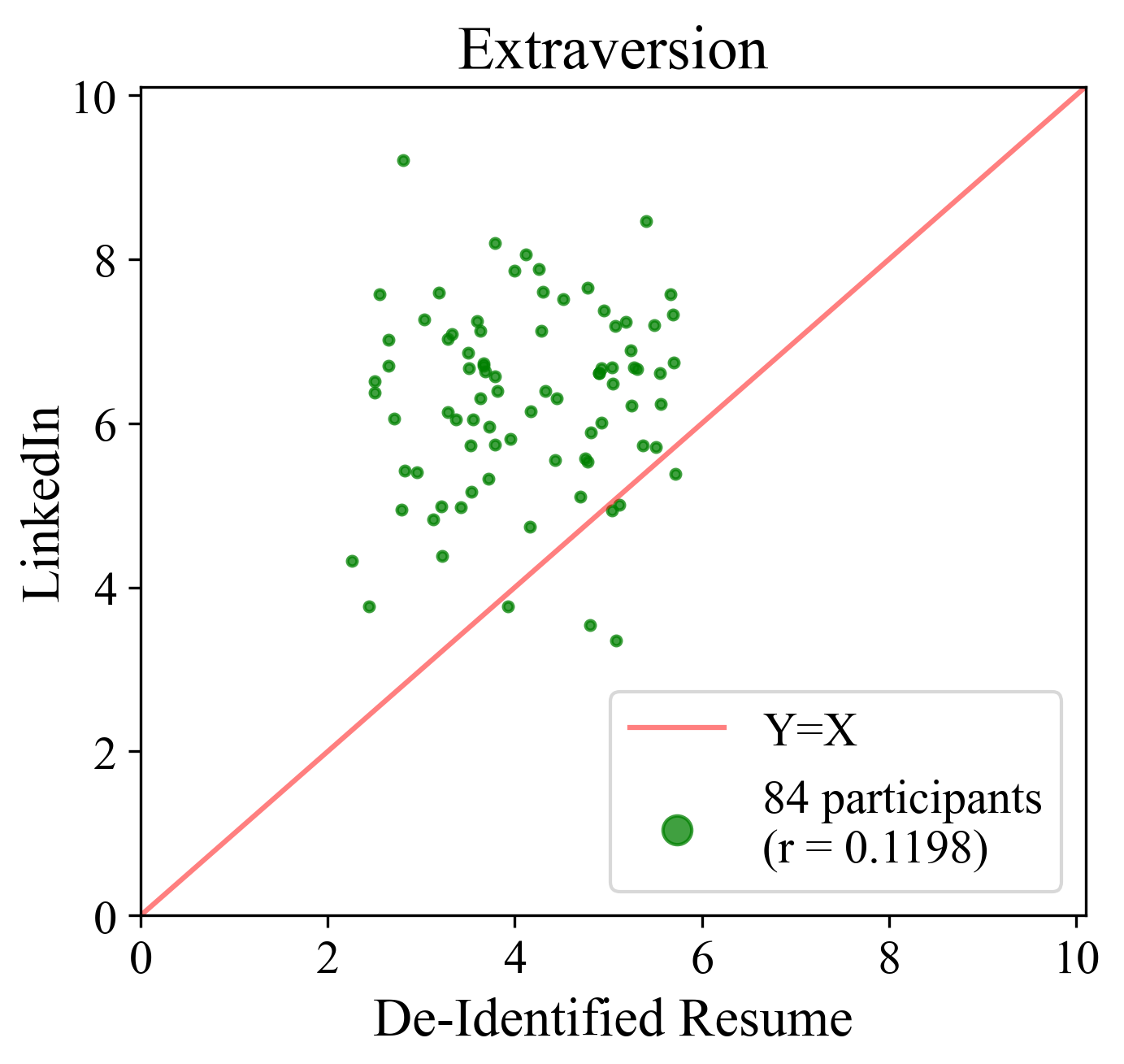}}
\caption{(a)~\humantic Dominance scores from de-identified and URL-embedded resumes. (b)~\humantic Extraversion scores produced by de-identified resumes and LinkedIn profiles.}
\label{fig:humantic_scatter}
\vspace{-0.5cm}
\end{figure}

\revv{We discovered substantial instability with regard to URL-embedding in resumes in \humantic.   Correlations between de-identified resumes and the same resumes with LinkedIn URLs embedded into them ranged from 0.077 (Extraversion) to 0.688 (Calculativeness).   We also discovered locational differences deemed significant by the Bonferroni threshold in Dominance, Steadiness, Big Five Conscientiousness, Extraversion, and Agreeableness. Under the more liberal Benjamini-Hochberg standard, there were also significant locational differences in DiSC Calculativeness and Openness.  Figure~\ref{fig:humantic_scatter}(a) gives a representative example; complete results are presented in Appendix~\ref{app:res:url}.}

We note that algorithm-time is unfortunately an unavoidable factor here; the two resume versions were run about four months apart. Furthermore, if we accept that \humantic uses information from LinkedIn profiles when it encounters embedded LinkedIn URLs, then we are also faced with a mismatch in participant-time.

\subsection{Source context} 
\label{sec:source_context_results}

\begin{figure}
\centering 
\includegraphics[width=0.9\textwidth]{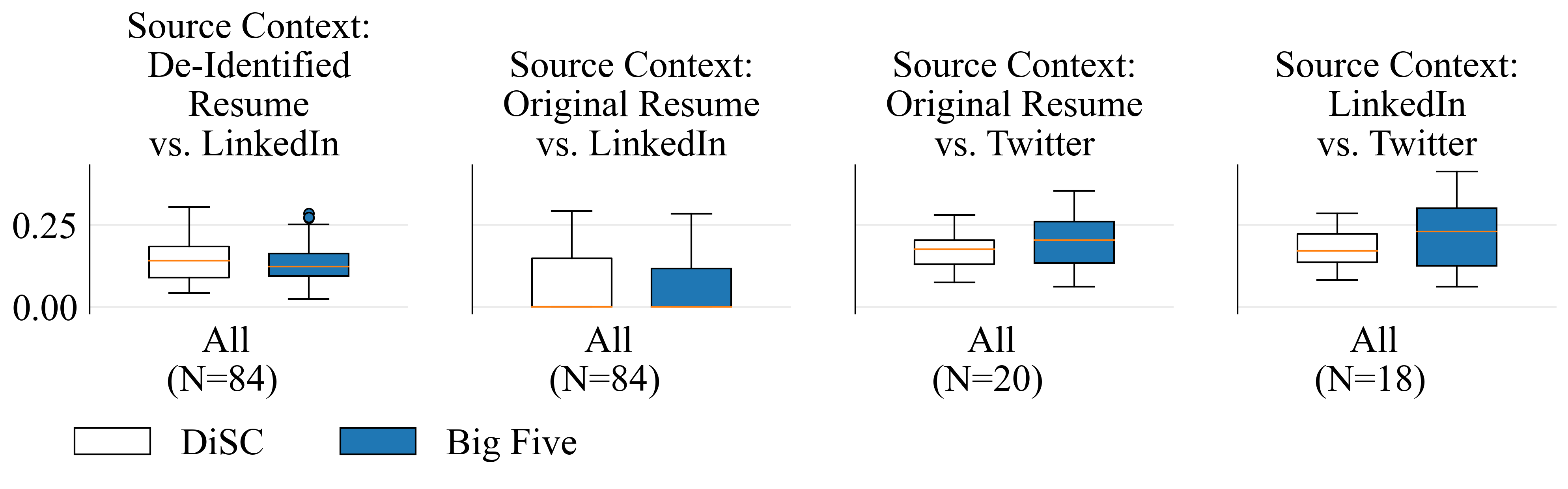}
\caption{Normalized L1 distances between \humantic DiSC and Big Five scores produced from pairs of treatments that vary with respect to their input source.}
\label{fig:humantic_box_source}
\end{figure}

\revv{\humantic and \crystal both displayed low stability across input sources. See Figure~\ref{fig:humantic_box_source} for comparison of L1 distances between each treatment of the input source facet in \humantic.}

\revv{\crystal's rank-order correlations between PDF resumes and LinkedIn profiles were all below the 0.90 threshold; they ranged from 0.233 (Dominance) to  0.526 (Influence). There was no significant locational instability in \crystal .
PDF resumes and LinkedIn URLs were scored the same day, and, as we will discuss in Section~\ref{sec:reproducibility_and_algorithm_time_results}, \crystal is immediately reproducible, and so we can rule out algorithm-time as a factor in this finding. Furthermore, for each candidate, this scoring took place within two weeks of resumes being submitted; thus, the participant-time of the resume matches very nearly to the participant-time of the LinkedIn. With all other facets being identical or near-identical, we can safely attribute the observed score differences to differences in source context.} 

\revv{De-identified resumes were submitted to \humantic 4 months after LinkedIn profiles had been run. This difference in algorithm-time hampers our interpretation of cross-profile correlations. Nonetheless, it is undeniably troublesome that the observed correlations are as low as 0.090 (Dominance), and that there were significant locational differences  under Bonferroni in Dominance and Extraversion, and under Benjamini-Hochberg in Steadiness and Openness. See Appendix~\ref{app:res:source_context} for details.}

\revv{We can avoid the issue of algorithm-time by using \humantic scores derived from original resumes, which were run at the same time as LinkedIn profiles. However, these results are somewhat misleading, as 57 of the 84 resumes in this experiment contained some form of LinkedIn URL. Considering the evidence that \humantic uses information directly from LinkedIn in such cases, correlations derived from original resumes are likely to overestimate cross-contextual stability. Nevertheless, the correlations we observe across all 84 participants range from 0.177 (Dominance) to 0.712 (Big Five Conscientiousness), with significant locational differences under Bonferroni in Dominance and Extraversion; and in Influence and Big Five Conscientiousness under Benjamini-Hochberg. We also found significant differences for non-native English speakers in Agreeableness under Benjamini-Hochberg. See Appendix~\ref{app:res:source_context} for details. Limiting analysis to the 27 participants whose original resumes contained no reference to LinkedIn, we find that the correlations straddle zero, ranging from -0.310 (Influence) to 0.297 (DiSC Calculativeness).}

\revv{Figure~\ref{fig:humantic_scatter}(b) highlights some of these results. Appendix~\ref{app:res:source_context} presents details of this experiment, and further includes a comparison of \humantic scores computed from Twitter to those computed from original resumes and from LinkedIn.}

\subsection{Algorithm-time}
\label{sec:reproducibility_and_algorithm_time_results}

\crystal results on resumes were reproducible immediately as well as one month later. We can conclude that \crystal's text prediction tool is deterministic and was not updated over the course of April 2021, when the experiment was performed.

\revv{\humantic results were not perfectly reproducible, even immediately. This may be explained by a non-deterministic prediction function, or by an online model that is updated with each prediction it makes. The latter explanation is in-line with our findings in the linkage investigations, where we observed that one call to the model can influence the outcome of other calls. Only Steadiness and DiSC Calculativeness remained constant for all participants when identical resumes were run back-to-back. One participant had changes in their Dominance and Influence scores (DiSC total normalized L1 difference was 0.005), and two participants had changes in their Big Five scores (maximum Big Five total normalized L1 difference was 0.003).  The correlations for immediate reproducibility were all above 0.95, and there were no significant locational differences.  }

\revv{After 31 days, rank-order correlations in \humantic ranged from 0.962 (Extraversion) to 0.998 (DiSC Calculativeness). Although the overall \humantic correlations across algorithm-time were all above the 0.95 threshold, we find that for non-native English speakers ($N=33$), Dominance ($r=0.946$) and Extraversion ($r=0.934$) both fell below 0.95. We also find significant instability in Openness under Benjamini-Hochberg.}

\revv{See Appendix~\ref{app:res:alg_time} for additional details about this experiment.}

\subsection{Participant-time}
\label{sec:participant_time_results} 

\begin{figure}
\centering 
\includegraphics[width=0.4\textwidth]{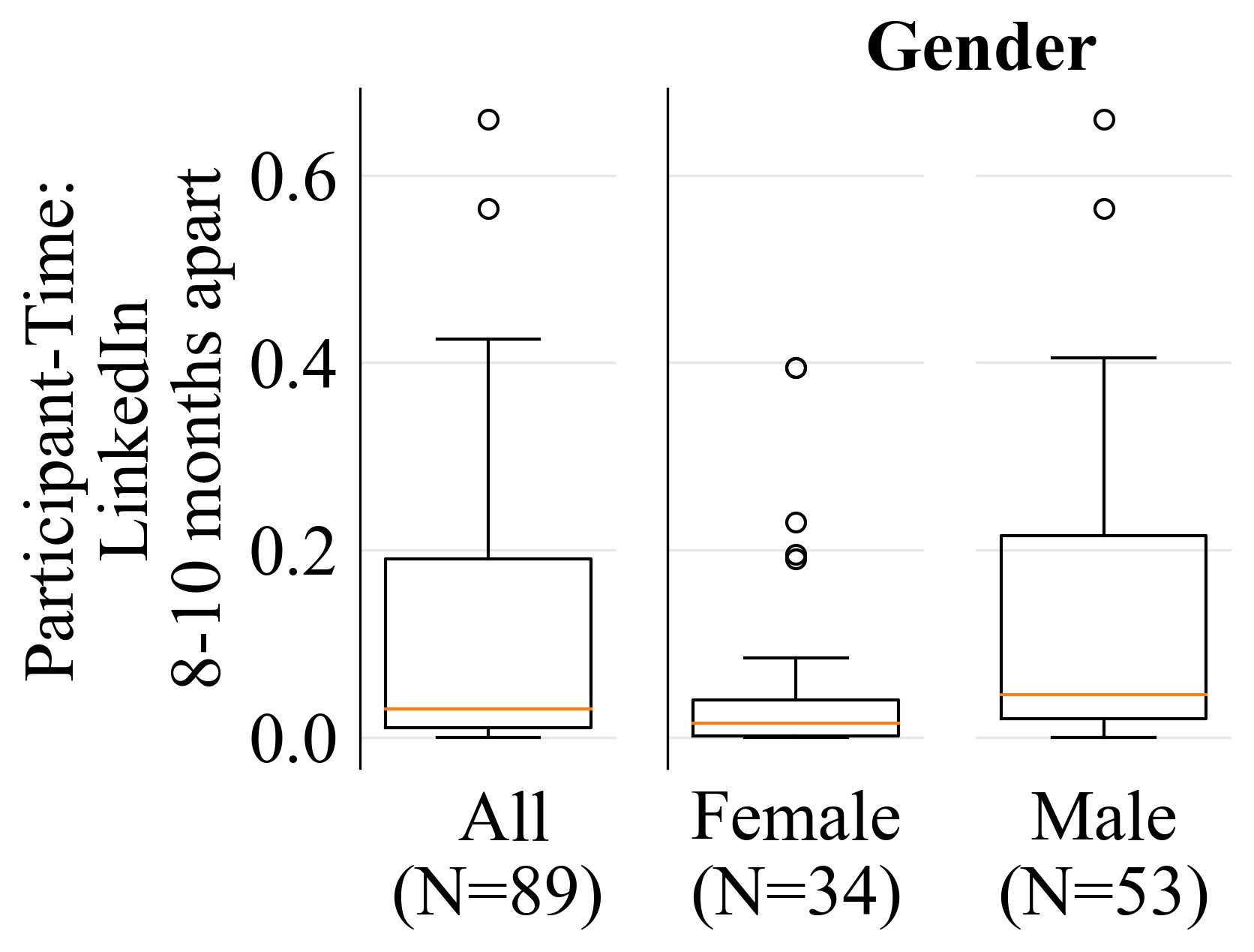}
\caption{Normalized L1 distances between \crystal DiSC scores produced from LinkedIn profiles scored 8-10 months apart.}
\label{fig:crystal_gender}
\end{figure}

\revv{\humantic scores on Twitter accounts showed no change over 7-9 months.  LinkedIn correlations across 7-9 months of participant-time were all below the 0.90 threshold: they ranged from 0.225 (Dominance) to 0.768 (Emotional Stability). Under Bonferroni correction, we found a significant difference in Big Five Conscientiousness scores, and under Benjamini-Hochberg we  found a significant difference in Agreeableness.}

\revv{\crystal LinkedIn correlations across 8-10 months of participant-time were all below the 0.90 threshold as well, ranging from 0.531 (Dominance) to 0.868 (Steadiness). We found that the reliability for male participants was particularly low ($N=53$, $r=0.232$). See Figure~\ref{fig:crystal_gender} for cross-gender comparison of L1 distances between participant-time treatments. There was no significant locational instability across participant-time in \crystal. See Appendix~\ref{app:res:part_time} for additional details about this experiment.}
\section{Discussion}
\label{sec:conc}

\subsection{Stability audit conclusions} 

\humantic and \crystal both exhibit low reliabilities across time and input source context. \humantic also exhibited low reliability with respect to the presence of LinkedIn URLs in resumes. \crystal's reliability with respect to resume format is unacceptably low as well. The correlations we observed \rev{allow} us to conclude that the tools cannot be considered valid instruments in high-stakes decisions.

Overall, each of these observed unreliabilities \rev{undermines} the cost and effort reduction that employers seek from candidate screening tools. Employers’ desire for valid decisions reflective of job performance is severely compromised by sensitivity to job-irrelevant factors. Thus, we find that \humantic’s sensitivities to participant-time, URL-embedding, and source context, and \crystal’s sensitivities to file format and source context, could be quite problematic for employers. The sensitivity of these algorithms to job-irrelevant factors is also a threat to individual fairness; a job seeker could reasonably conclude from the present audit that \humantic and \crystal are both likely to judge their job-worthiness unfairly, letting meaningless criteria dictate their outcomes.

These unreliabilities are also at odds with the trustworthiness that society seeks in its AI products. 
\humantic's lack of reproducibility is a particularly insidious violation of trustworthiness, because it undermines the power of audits on its system. Although \humantic's stability over algorithm-time exceeds Nunnally and Bernstein's~\cite{nunnally_psychometric_1994} classical 0.95 reliability threshold for tests used to make decisions about individuals (see Section~\ref{sec:psychometric_theory}), the Supreme Audit Institutions of Finland, Germany, the Netherlands, Norway and the UK~\cite{supreme} have asserted that reproducibility is ``a mandatory condition for reliability.'' Irreproducibility resulting from frequently or continuously updated models poses a threat to the ongoing monitoring and auditing necessary to ensure a system is working as expected~\cite{bogen_help_2018,koshiyama_towards_2021}. 

Finally, \humantic's and \crystal's lack of transparency regarding training data and model architecture are at odds with privacy concerns. \humantic's deceptive and ineffective opt-out option is an example of what Ajunwa~\cite{ajunwa_auditing_2021} calls ``algorithmic blackballing,'' whereby an applicant’s profile is allowed to live on past its shelf-life. This is especially dangerous in combination with the potential to leverage \humantic's email linkage mechanism in an adversarial attack. \humantic's failed opt-out \rev{option may} also violate the California Consumer Privacy Act's right to delete~\cite{ca_privacy}.

\subsection{Study limitations} 
\label{sec:study_limitations}

In our audit we do not conduct stakeholder evaluations.  Several audits and framework documents emphasize the importance of algorithmic impact assessment and stakeholder evaluations~\cite{brown_algorithm_2021,fjeld_principled_2020,ORCAA,raji_closing_2020,razavi_future_2021,sloane_silicon_2021}. 
Metcalf \etal~\cite{metcalf_algorithmic_2021} explain that an external audit must not stand in as an impact assessment. Without collaboration of internal agents, third parties do not have access to design decisions or stakeholder interviews, and cannot directly influence change in the design or operation of the algorithm should it be needed. Per Ajunwa~\cite{ajunwa_auditing_2021}, algorithms need to be audited internally as well as externally.

\revv{
Although this audit considers various dimensions of reliability and stability, the analysis is not comprehensive. We have constrained our audit methodology to analyze the numerical scores produced by personality prediction AI that claim to offer a quantitative measure of ``personality'' (such as the DiSC and Big Five scores produced by \crystal and \humantic). However, much of the advertising of such tools focus on the profiles holistically, not just on the scores. Further \crystal and \humantic both categorize candidates into one of several types and produce descriptive personality profiles. Written profiles are likely influential in hiring decisions, however, in the interest of keeping the scope of our work manageable, we leave a treatment of stability in these textual profiles to future work.}

\revv{
The audit methodology is also limited by its emphasis on comparing pairs of control and treatment scores. For example: \humantic often fails to produce profiles from inputs (see the discrepancies between number of inputs submitted and number of profiles produced in Table~\ref{humantic_runs}). This is especially common when using Twitter profiles. By simply disregarding the failed inputs, we may be introducing some sampling bias into our results. Furthermore, such non-results may exhibit problematic biases~\cite{ORCAA}.
}


Our study population was constrained to technical graduate students at NYU, studying in the realms of computer and data science. This was done in an attempt to control for differences in algorithm response due to characteristics such as job field, experience level, and writing style. We also felt that this restriction more closely replicated a pool of candidates who might realistically be compared to one another in a job search. However, this narrowness, and our modest cohort size ($N=94$), restrict the generalizability of the results of our audit of \crystal and \humantic.

Additionally, this audit evaluates only the intermediate personality profile results, and does not relate them to hiring outcomes. Our audit did not use the ``job fit'' or ``match score'' features because, as external auditors, we did not have access to information on how ideal candidates are defined or how thresholds are set. Without this information, we cannot assess outcomes-based fairness metrics. This means that critical questions of discrimination remain out of scope for this study. We caution that the adverse impact of human-in-the-loop hiring systems must be assessed on an employer-by-employer basis in order to account for crucial implementation \rev{details and} differences in the context of use.

\section{Conclusions and Future Work}
\label{sec:outlook}

\revv{In this paper, we investigated the reliability of algorithmic personality tests used in hiring. We gave an overview of the key literature on psychometric testing applied to hiring and in algorithm auditing, and found that, although reliability is seen as a necessary condition for the validity of a psychometric instrument, it has not received substantial treatment in algorithm audits.  Based on this observation, we developed a socio-technical audit methodology, informed by psychometric theory and sociology, to test the stability of black-box algorithms that predict personality for use in hiring. We also developed an open-source software library to automates the quantitative components of this framework. We then instantiated this methodology in an external audit of two systems, \humantic and \crystal, using a dataset of job applicant profiles collected through an IRB-approved study. \revv{Using our audit methodology, we found that} that both systems lack reliability across key facets of measurement, and concluded that they cannot be considered valid personality assessment instruments.}

The present study demonstrates that stability, though often overlooked in algorithm audits, is an accessible metric for external auditors. We found that stability is highly relevant to the application of personality prediction. Furthermore, because reliability is a prerequisite of validity, stability is in fact relevant whenever validity is.  Importantly, we  note that, \rev{while reliability} is a necessary condition for validity, it is not a sufficient condition.  Further evidence of domain-specific validity is essential to support the  use of algorithmic personality tests in hiring.

Our methodology can be used by employers to make informed purchasing and usage decisions, and to better interpret algorithm outputs, by legislators to guide regulation, and by consumers to make informed decisions about how and when to disclose their information to potential employers. \revv{Our open-source software library reduces the amount of effort that would be required to conduct such analyses.}

\revv{Moreover, given its modular design, our software library can be easily extended to support other reliability-related measures. As mentioned in Section \ref{sec:methodology}, we envision an extension of the \emph{Treatment Generator} that can (at least partially) automate the creation of treatments, for example, by sampling values for a particular feature from an appropriate distribution, or by automatically perturbing values in text features. The library's visualization capabilities, which currently include scatterplots and boxplots, can also be extended to facilitate the generation of audit results that are amenable to a wide variety of stakeholders, both technical and non-technical. The library already computes statistics broken down by demographic groups of interest, and can also easily be extended to compute fairness-related measures.}

Algorithmic audits must not be one-size-fits-all. The tendency of auditors, especially within the hiring domain, to rely on legal frameworks as a scoping mechanism is likely to leave important risks undetected. Current legal frameworks are insufficient; furthermore, legality does not equate to ethics. Instead, we recommend that auditors interrogate the assumptions operationalized by systems, and design audits accordingly. 

Finally, we note that this work was conducted by an interdisciplinary team that included computer and data scientists, a sociologist, an industrial psychologist, and an investigative journalist.  This collaboration was both necessary and challenging, requiring us to reconcile our approaches and methodological toolkits, forging new methods for interdisciplinary collaboration. 

\begin{acknowledgements}
We thank Dhara Mungra for her work on data collection and preliminary analysis, and Daphna Harel and Joshua Loftus for their advice on statistical methods.
\end{acknowledgements}
\section{Declarations}

\subsection{Funding} 
This research is supported in part by NSF Awards No. 1934464, 1922658, and 1916505 (PI Stoyanovich), and by the NYU Center for the Humanities Digital Humanities Seed Grant (PIs  Schellmann and Sloane). 

\subsection{Conflict of interest}
The authors declare that they have no conflict of interest.

\subsection{Availability of data and material}
Anonymized datasets generated for and analysed during the current study are available at \url{https://github.com/DataResponsibly/hiring-stability-audit/tree/main/data}.

\subsection{Code availability}
Code for data cleaning and analysis is provided for replication. It is available at \url{https://github.com/DataResponsibly/hiring-stability-audit}.

\subsection{Ethics approval}
All procedures performed involving human participants were in accordance with the ethical standards of the NYU institutional review board and with the 1964 Helsinki declaration and its later amendments or comparable ethical standards.

\subsection{Consent to participate}
Informed consent was obtained from all individual participants included in the study.

\subsection{Consent for publication}
Participants granted informed consent to publish information not containing identifiers, including personality profile results and demographic data.

\bibliographystyle{spmpsci}       
\bibliography{AI_Hiring_Audit}    

\pagebreak

\appendix 
\section{Accounting for multiple hypothesis testing}
\label{app:stats}
\paragraph{Bonferroni correction} This method controls the family-wise error rate. It is guaranteed to falsely reject the null hypothesis no more often than the nominal significance level, however, it can be overly conservative, especially when sample sizes are low (\ie it can falsely accept the null hypothesis more often than the nominal significance level implies)~\cite{bonferroni}.

$$\alpha_{\text{Bonferroni}} = \frac{\alpha_{\text{nominal}}}{\text{\# tests performed}}$$
        
\paragraph{Benjamini-Hochberg correction} This is a less conservative approach that controls the false discovery rate. The procedure ranks obtained p-values in ascending order and uses these ranks to derive corrected thresholds, which range between $\alpha_{\text{Bonferroni}}$ and $\alpha_{\text{nominal}}$~\cite{benjamini_hochberg}.

$$\alpha_{\text{Benjamini-Hochberg}} = \frac{\text{p-value rank}}{\text{\# tests performed}}\alpha_{\text{nominal}}$$

\section{Additional audit details}
\label{app:details}

\subsection{Participant demographics}
\label{app:details:demographics}

Table~\ref{demographics} supplements the description of \emph{primary data collection} discussed in Section~\ref{sec:inputs_and_experiments} with information about participant demographics.  Additional information about how participants were recruited is available but cannot be disclosed at this time due to double-blind reviewing requirements.

\begin{table}[h!]
\centering
\small 
\caption{Demographics of IRB-approved user study participants.}
\label{demographics}
    \begin{tabular}{p{0.2\textwidth} | p{0.08\textwidth} | p{0.08\textwidth}}
    \hline\noalign{\smallskip}
    Group & N & \%  \\
    \noalign{\smallskip}\hline\noalign{\smallskip}
    All & 94 & 100\% \\
    \rowcolor[HTML]{C0C0C0} Gender &  &  \\
    Male & 56 & 60\% \\
    Female & 36 & 38\% \\
    Other & 2 & 2\% \\
    \rowcolor[HTML]{C0C0C0} Race &  &  \\
    Asian & 57 & 61\% \\
    White & 24 & 26\% \\
    Other & 12 & 13\% \\
    No Answer & 1 & 1\% \\
    \noalign{\smallskip}\hline
    \end{tabular}
    \hfill
    \begin{tabular}{p{0.2\textwidth} | p{0.08\textwidth} | p{0.08\textwidth}}
    \hline\noalign{\smallskip}
    Group & N & \%  \\
    \noalign{\smallskip}\hline\noalign{\smallskip}
    \rowcolor[HTML]{C0C0C0} Birth Country &  &  \\
    India & 34 & 36\% \\
    USA & 28 & 30\% \\
    China & 12 & 13\% \\
    Other & 18 & 19\% \\
    No Answer & 2 & 2\% \\
    \rowcolor[HTML]{C0C0C0} Primary Language &  &  \\
    English & 60 & 64\% \\
    Other & 34 & 36\% \\
    \noalign{\smallskip}\hline
    \end{tabular}
\end{table}

\subsection{Treatments for each facet}
\label{app:details:calls}
\begin{table*}[t!]
\small 
\centering
\caption{Details of \humantic runs (\ie sets of score-generating calls to \humantic models).}
\label{humantic_runs}
\begin{tabular}{p{0.15\textwidth} | p{0.07\textwidth} | p{0.2\textwidth} | p{0.07\textwidth} | p{0.07\textwidth}} 
\hline\noalign{\smallskip}
Input & Run ID & \humantic Run Dates & Inputs Submitted & Profiles Produced  \\
\noalign{\smallskip}\hline\noalign{\smallskip}
Original Resume & HRo1 & 11/23/2020 - 01/14/2021 & 89 & \rev{88} \\
De-Identified Resume & HRi1 & 03/20/2021 - 03/28/2021 & 89 & 89 \\
De-Identified Resume & HRi2 & 04/20/2021 - 04/28/2021 & 89 & 89 \\
De-Identified Resume & HRi3 & 04/20/2021 - 04/28/2021 & 89 & 89 \\
DOCX Resume & HRd1 & 03/20/2021 - 03/28/2021 & 89 & 89 \\
URL-Embedded Resume & HRu1 & 04/09/2021 - 04/11/2021 & \rev{86} & \rev{86} \\
LinkedIn & HL1 & 11/23/2020 - 01/14/2021 & 92 & 88 \\
LinkedIn & HL2 & 08/10/2021 - 08/11/2021 & 92 & \rev{91} \\
Twitter & HT1 & 11/23/2020 - 01/14/2021 & 32 & 21 \\
Twitter & HT2 & 08/10/2021 - 08/11/2021 & 32 & 21 \\
\noalign{\smallskip}\hline
\end{tabular}
\end{table*}

\begin{table*}
\small 
\caption{Details of \crystal runs (\ie sets of score-generating calls to \crystal models).}
\label{crystal_runs}
\begin{tabular}{p{0.25\textwidth} | p{0.1\textwidth} | p{0.25\textwidth} | p{0.1\textwidth} | p{0.1\textwidth}} 
\hline\noalign{\smallskip}
Input & Run ID & \crystal Run Dates & Inputs Submitted & Profiles Produced  \\
\noalign{\smallskip}\hline\noalign{\smallskip}
Raw Text Resume & CRr1 & 03/31/2021 - 04/02/2021 & 89 & 89 \\
Raw Text Resume & CRr2 & 05/01/2021 - 05/03/2021 & 89 & 89 \\
Raw Text Resume & CRr3 & 05/01/2021 - 05/03/2021 & 89 & 89 \\
PDF Resume & CRp1 & 11/23/2020 - 01/14/2021 & 89 & 89 \\
LinkedIn & CL1 & 11/23/2020 - 01/14/2021 & 92 & 91 \\
LinkedIn & CL2 & 09/13/2021 - 09/16/2021 & 89 & 89 \\
\noalign{\smallskip}\hline
\end{tabular}
\end{table*}

Details of score-generating model calls to generate treatments for each facet, discussed in Section~\ref{sec:inputs_and_experiments}, are presented in Table~\ref{humantic_runs} for \humantic and in Table~\ref{crystal_runs} for \crystal. In these tables, we list the type of input (\eg Original Resume or LinkedIn profile), the identifier of the run that corresponds to this input, and the range of dates over which the system (\humantic or \crystal) was fed this type of input.  We also list input size (``Inputs Submitted'') and output size (``Profiles Produced'').  Note that output size may be smaller compared to input size, and sometimes substantially so.  For example, for runs HT1 and HT2, we used 32 Twitter handles as input to \humantic, but we took only 21 personality profiles produced as output into consideration.  This is because \humantic did not produce personality profiles from the remaining 11 accounts, but instead returned errors saying the Twitter profiles were ``thin.''

\subsection{Choice of stability metrics}
\label{app:details:metrics}
This section describes the metrics used to assess facet-specific stability.
\begin{itemize}[\noindent]
    \item[{\bf Rank-order stability.}] Because DiSC scores were discontinuous in \crystal, we use Spearman rank correlation rather than Pearson's correlation coefficient to quantify rank-order stability.  Rank\rev{-}order stability results are presented in Tables~\ref{crystal_corr_disc}, \ref{humantic_corr_disc}, and \ref{humantic_corr_big5}. 
    \item[{\bf Locational stability.}] Similarly, we use the Wilcoxon signed\rev{-}rank test to assess the significance of paired differences. Unlike the Student's t-test, the Wilcoxon signed-rank test does not assume the data is normally distributed. Locational stability results can be found in Tables~\ref{crystal_wc_disc}, \ref{humantic_wc_disc}, and \ref{humantic_wc_big5}. \rev{We start with a nominal $\alpha$ of 0.05. In \crystal, we test the median change of the four DiSC traits across five facets, for a total of 20 tests and a Bonferroni-corrected $\alpha$ of 0.0025.  In \humantic, we test the Big Five traits and the four DiSC traits across eleven facets, for a total of 99 tests and a Bonferroni-corrected $\alpha$ of $5.05 \times 10^{-4}$.}
    \item[{\bf Total change.}] To compute total change, we calculate the L1 distance between the output vectors of the two runs for each subject. In order to compare results across different scales, this distance is normalized by the total range of output space. The normalization constant is the inverse of the sum of possible score ranges for each trait in the category. For example, \humantic produces four DiSC scores each measured on a scale from 0 to 10, so we divide the DiSC L1 distances by 40. Because \crystal constrains their DiSC scores to sum to 100, the maximum possible L1 change is 200, and we therefore use a normalization constant of 200.
    \item[{\bf Subgroup stability.}] We use demographic information provided in our survey to estimate rank-order stability, locational stability, and normalized L1 distance within subgroups defined by gender and primary language. With only 94 participants, we lacked the statistical power to perform statistical analysis on the smaller subgroups (e.g. birth country, race). 

\end{itemize}

\section{Additional results}
\label{app:res}



\subsection{Inclusion of LinkedIn URL in resume}
\label{app:res:url}


%
We discovered \rev{locational differences deemed significant by the Bonferroni threshold in} Dominance (de-identified median 6.90, URL-embedded median 5.65; Wilcoxon $p<10^{-6}$), Big Five Conscientiousness (de-identified median \rev{5.60}, URL-embedded median 6.17; Wilcoxon $p=2.1 \times 10^{-5}$), and Extraversion (de-identified median 4.14, URL-embedded median \rev{6.38}; Wilcoxon $p<10^{-6}$). \rev{Under the more liberal Benjamini-Hochberg standard, there were also significant locational differences in DiSC Calculativeness (de-identified median 7.50, URL-embedded median 8.00; Wilcoxon $p=4.7 \times 10^{-3}$),  Openness (de-identified median 6.14, URL-embedded median 5.90; Wilcoxon $p=2.5 \times 10^{-3}$)}, Steadiness (de-identified median 5.00, URL-embedded median 5.60; Wilcoxon $p=4.8 \times 10^{-4}$), and Agreeableness (de-identified median 5.56, URL-embedded median \rev{6.07}; Wilcoxon $p=1.6 \times 10^{-4}$).


Correlations between scores derived from LinkedIn profiles and from URL-embedded resumes ranged from 0.156 (Dominance) to 0.702 (Emotional Stability), and there was a significant difference in the medians of Big Five Conscientiousness (LinkedIn 5.72, resume 6.19; Wilcoxon $p=4.3 \times 10^{-5}$)\rev{, per the Bonferroni-adjusted threshold}. \rev{Under Benjamini-Hochberg correction, the differences in Dominance (LinkedIn median 4.90, resume median 5.60; Wilcoxon $p=6.6 \times 10^{-3}$) and Agreeableness (LinkedIn median 5.81, resume median 6.06; Wilcoxon $p=6.8 \times 10^{-3}$) were significant as well.} We predicted higher correlations under the embedding hypothesis, but a four month gap in algorithm-time as well as participant-time is likely to degrade the correlations significantly. Still, LinkedIn scores are more highly correlated with URL-embedded resumes than they are with de-identified resumes.  Although instability due to algorithm\rev{-}time is not guaranteed to increase monotonically with chronological time, this finding holds slightly more weight given that there \rev{were} two more weeks of time between the LinkedIn and URL-embedding resume scoring. We also find that scores from URL-embedded resumes correlate slightly better with those from LinkedIn (generated four months earlier) than they do with those from de-identified resumes (generated just 2 weeks earlier).

\subsection{Source context}
\label{app:res:source_context}


Comparing de-identified resumes to LinkedIn profiles in \humantic, we found significant locational differences \rev{under Bonferroni} in Dominance (LinkedIn median 4.85, resume median 6.85; Wilcoxon $p<10^{-6}$) and Extraversion (LinkedIn median 6.44, resume median 4.06; \rev{Wilcoxon }$p<10^{-6}$)\rev{, and under Benjamini-Hochberg in Steadiness (LinkedIn median 5.30, resume median 5.00; Wilcoxon $p=1.3 \times 10^{-3}$) and Openness (LinkedIn median 6.01, resume median 6.14; Wilcoxon $p=7.7 \times 10^{-3}$)}.

When original resumes were compared to LinkedIn profiles in \humantic, we observed significant locational differences \rev{under Bonferroni} in Dominance (LinkedIn median 4.85, resume median 5.95; Wilcoxon $p= 7x10^{-6}$) and Extraversion (LinkedIn median 6.44, resume median 5.75; Wilcoxon $p= 6.9x10^{-5}$) \rev{, and significant locational differences under Benjamini-Hochberg in Influence (LinkedIn meidan 4.60, resume median 4.85; Wilcoxon $p=5.0 \times 10^{-3}$)} and Big Five Conscientiousness (LinkedIn median 5.73, resume median 5.98; Wilcoxon $p= 2.8x10^{-4}$). Although there was not any significant locational instability for Agreeableness overall, we found that for non-native English speakers, the median Agreeableness score on resumes (5.99) was significantly different under Benjamini-Hochberg ($p=6.1 \times 10^{-3}$) from the median score on LinkedIn (5.63).

Comparing \humantic scores from Twitter to those from original resumes, we find correlations ranging from -0.521 (Dominance) to 0.232 (Big Five Conscientiousness). We easily avoid the issue of algorithm-time by using original resumes, which were run the same day as Twitter. None of the original resumes contain references to participants' Twitter accounts, \rev{and furthermore we did not find evidence of linkage with Twitter profiles, }so we need not worry about data leakage in this case. A major caveat to this result is the small sample size ($N=20$). Although the locational differences were insignificant when compared to the Bonferroni-corrected threshold, the Benjamini-Hochberg correction found significant locational differences in Agreeableness (resume median 6.37, Twitter median 3.32; Wilcoxon $p=2.0 \times 10^{-3}$) and Emotional Stability (resume median 5.42, Twitter median 7.97; Wilcoxon $p=1.0 \times 10^{-3}$).  Although there was not any significant locational instability for Openness overall, we found that for male participants, the median Openness score on resumes (5.71) was significantly different under Benjamini-Hochberg ($p=6.1 \times 10^{-3}$) from the median score on Twitter (8.50).

Finally, we compare the \humantic scores from LinkedIn and Twitter. Again we have a small sample size ($N=18$), however the results are striking. Only one of the correlations is positive (Influence, $r=0.020$), and the others are as low as -0.433 (DiSC Calculativeness). Again there are no significant locational differences \rev{under Bonferroni, but using the Benjamini-Hochberg correction we find significant differences in Openness (LinkedIn median 5.82, Twitter median 8.16; Wilcoxon $p=2.3 \times 10^{-3}$), Big Five Conscientiousness (LinkedIn median 5.77, Twitter median 7.16; Wilcoxon $p=4.7 \times 10^{-3}$), Extraversion (LinkedIn median 6.80, Twitter median 4.72; Wilcoxon $p=6.7 \times 10^{-4}$), Agreeableness (LinkedIn median 6.32, Twitter median 3.32; Wilcoxon $p=4.7 \times 10^{-3}$), and Emotional Stability (LinkedIn median 4.86, Twitter median 7.97; Wilcoxon $p=6.7 \times 10^{-4}$)}. Although there was not any significant locational instability for Dominance overall, we found that for male participants, the median Dominance score on LinkedIn (4.30) was significantly different under Benjamini-Hochberg ($p=2.0 \times 10^{-3}$) from the median score on Twitter (6.90). Participant-time and algorithm-time are both guaranteed to be constant in this experiment, as profiles were generated on the same day.

Complete experimental results for \humantic are listed in Tables~\ref{humantic_corr_disc}, \ref{humantic_corr_big5}, \ref{humantic_wc_disc}, and \ref{humantic_wc_big5}.

\subsection{Algorithm time}
\label{app:res:alg_time}

\begin{figure}
\centering 
\includegraphics[width=0.6\textwidth]{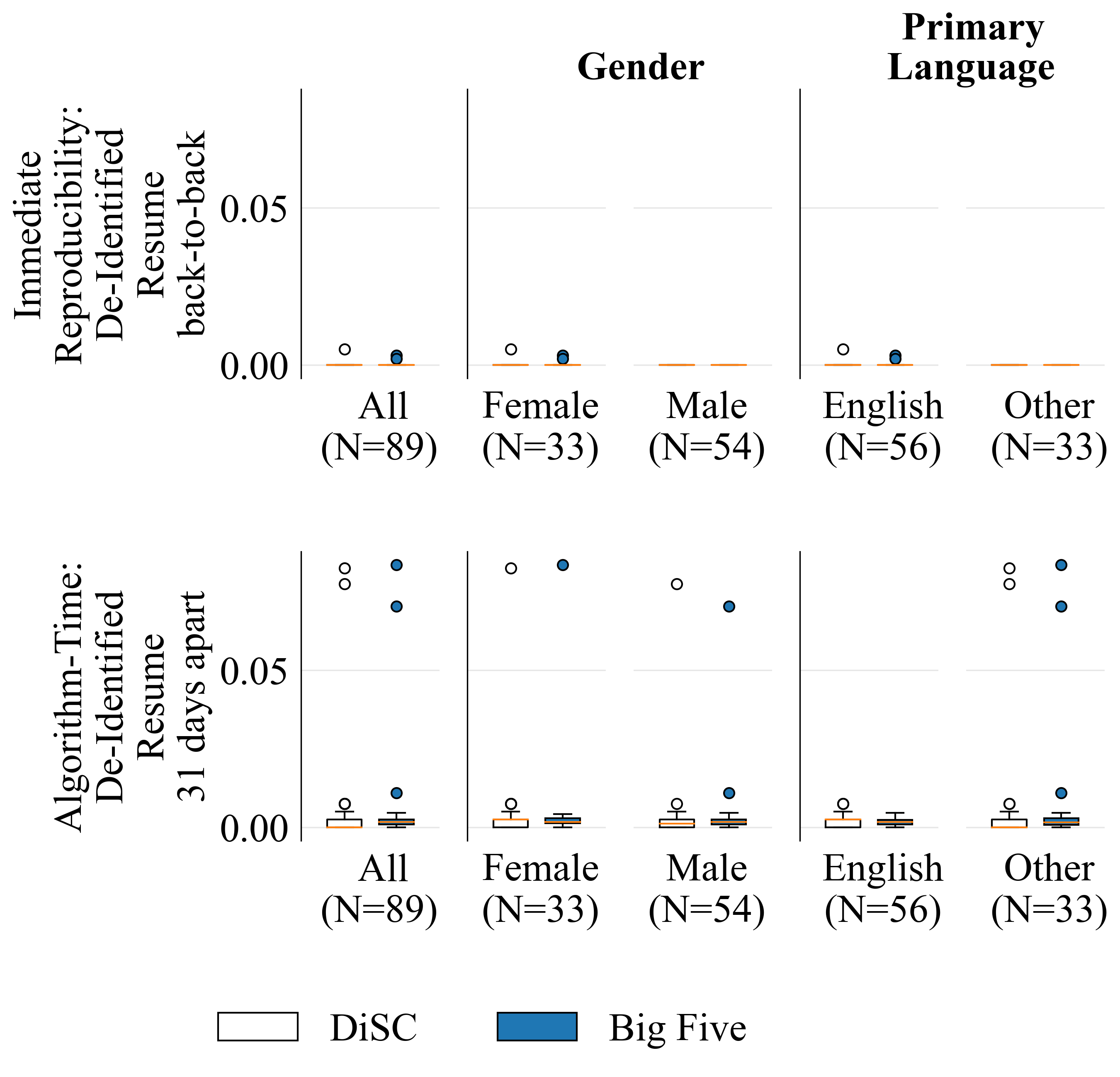}
\caption{Normalized L1 distances between \humantic DiSC and Big Five scores produced from identical resumes scored at different points in time.}
\label{fig:humantic_box_atime}
\end{figure}

Figure~\ref{fig:humantic_box_atime} shows that substandard sub-group correlations result from two participants whose resumes were scored very differently by \humantic a month apart; we also note that the lack of immediate reproducibility we observed in \humantic did not affect these two particular individuals. \rev{We did not find any significant locational differences across algorithm-time using the Bonferroni correction, but under Benjamini-Hochberg we found significant differences in Openness, where the median decreased from 6.15 to 6.13 over the course of a month (Wilcoxon $p=7.1 \times 10^{-3}$).}

\subsection{Participant time}
\label{app:res:part_time}

Built into the substandard correlations across participant-time in \humantic LinkedIn runs is the corrosive effect of 7-9 months of participant-time; this helps to explain, but does not justify, the unacceptably low test-retest reliability. 

Under Bonferroni correction, we found the following significant difference in \humantic LinkedIn across 7-9 months of participant-time: Big Five Conscientiousness scores, with the median increasing from 5.72 to 6.17 (Wilcoxon $p=4 \times 10^{-6}$). Under Benjamini-Hochberg we also found a significant difference in Agreeableness, where the median increased from 5.81 to 5.99 (Wilcoxon $p=7.2 \times 10^{-3}$). Complete experimental results for \humantic are listed in Tables~\ref{humantic_corr_disc}, \ref{humantic_corr_big5}, \ref{humantic_wc_disc}, and \ref{humantic_wc_big5}.

\begin{table*}
\caption{Rank-order stability of \crystal DiSC scores, as measured by Spearman's rank correlations. Reliabilities below 0.90 highlighted in yellow; those between 0.90 and 0.95 highlighted in lighter yellow. Results are discussed in Sections~\ref{sec:file_format_results}, \ref{sec:url_embedding_results}, \ref{sec:source_context_results}, \ref{sec:reproducibility_and_algorithm_time_results}, and \ref{sec:participant_time_results}.}
\label{crystal_corr_disc}
\begin{tabular}{>{\raggedright}p{0.14\textwidth} |>{\raggedright}p{0.2\textwidth} | >{\centering}p{0.02\textwidth} | >{\centering}p{0.1\textwidth} | >{\centering}p{0.08\textwidth} | >{\centering}p{0.09\textwidth} | >{\centering\arraybackslash}p{0.13\textwidth}} 
\hline\noalign{\smallskip} Facet & Input Versions &   N &  Dominance &  Influence &  Steadiness &  Conscientious-ness \\
\noalign{\smallskip}\hline\noalign{\smallskip}
File Format &  Raw Text Resume (CRr1) vs. PDF Resume (CRp1) &  89 &    \signifmed 0.8225 &    \signifmed 0.8260 &    \signiflow  0.9184 &         \signiflow  0.9114 \\
Source Context &         PDF Resume (CRp1) vs. LinkedIn (CL1) &  86 &   \signifmedhigh  0.2335 &   \signifmed  0.5258 &     \signifmed 0.5103 &          \signifmedhigh 0.3585 \\
Immediate Reproducibility &    Raw Text Resume back-to-back (CRr2 vs. CRr3) &  89 &     1.0000 &     1.0000 &      1.0000 &           1.0000 \\
Algorithm-Time &   Raw Text Resume 31 days apart (CRr1 vs. CRr2) &  89 &     1.0000 &     1.0000 &      1.0000 &           1.0000 \\
 Participant-Time &      LinkedIn 8-10 months apart (CL1 vs. CL2) &  89 &     \signifmed 0.5314 &    \signifmed 0.7062 &     \signifmedlow 0.8676 &         \signifmed  0.7811 \\
 \noalign{\smallskip}\hline
\end{tabular}
\end{table*}

\begin{table*}
\small 
\caption{Rank-order stability of \humantic DiSC scores, as measured by Spearman's rank correlations. Reliabilities below 0.90 highlighted in yellow. Results are discussed in Sections~\ref{sec:file_format_results}, \ref{sec:url_embedding_results}, \ref{sec:source_context_results}, \ref{sec:reproducibility_and_algorithm_time_results}, and \ref{sec:participant_time_results}.}
\label{humantic_corr_disc}
\begin{tabular}{>{\raggedright}p{0.14\textwidth} | >{\raggedright}p{0.2\textwidth} | >{\centering}p{0.02\textwidth} | >{\centering}p{0.1\textwidth} | >{\centering}p{0.08\textwidth} | >{\centering}p{0.09\textwidth} | >{\centering\arraybackslash}p{0.13\textwidth}} 
\hline\noalign{\smallskip}
Facet & Input Versions &   N &  Dominance &  Influence &  Steadiness &  Calculativeness \\
\noalign{\smallskip}\hline\noalign{\smallskip}
File Format &          De-Identified Resume (HRi1) vs. DOCX Resume (HRd1) &  89 &     0.9956 &     0.9924 &      0.9978 &           0.9959 \\
    URL Embedding &  URL-Embedded Resume (HRu1) vs. De-Identified Resume (HRi1) &  86 &   \signifmedhigh  0.3570 &   \signifmed  0.6253 &      \signifmed 0.5480 &          \signifmed 0.6878 \\
    URL Embedding &     URL-Embedded Resume (HRu1) vs. LinkedIn (HL1) &  83 &   \signifmedhigh  0.1555 &    \signifmedhigh 0.3382 &      \signifmed 0.6074 &          \signifmedhigh 0.4701 \\
   Source Context &            De-Identified Resume (HRi1) vs. LinkedIn (HL1) &  84 &   \signifmedhigh  0.0903 &    \signifmedhigh 0.2553 &   \signifmedhigh   0.3941 &        \signifmedhigh   0.3331 \\
   Source Context &                  Original Resume (HRo1) vs. LinkedIn (HL1) &  84 &   \signifmedhigh  0.1775 &  \signifmedhigh   0.4016 &      \signifmed 0.6939 &         \signifmed  0.6249 \\
   Source Context &                   Original Resume (HRo1) vs. Twitter (HT1) &  20 &   \signifhigh -0.5211 &    \signifmedhigh 0.1026 &      \signifmedhigh 0.0382 &         \signifhigh -0.1475 \\
   Source Context &                          LinkedIn (HL1) vs. Twitter (HT1) &  18 &   \signifhigh -0.1317 &    \signifmedhigh 0.0203 &     \signifhigh -0.1120 &        \signifhigh  -0.4329 \\
  Immediate Reproducibility &            De-Identified Resume back-to-back (HRi2 vs. HRi3) &  89 &     0.9999 &     1.0000 &      1.0000 &           1.0000 \\
   Algorithm-Time &            De-Identified Resume 31 days apart (HRi1 vs. HRi2) &  89 &     0.9726 &     0.9948 &      0.9925 &           0.9980 \\
 Participant-Time &                     LinkedIn 7-9 months apart (HL1 vs. HL2) &  88 &     \signifmedhigh 0.2248 &     \signifmedhigh 0.4186 &    \signifmed  0.6597 &         \signifmed  0.5827 \\
 Participant-Time &                      Twitter 7-9 months apart (HT1 vs. HT2) &  21 &     1.0000 &     1.0000 &      1.0000 &           1.0000 \\
 \noalign{\smallskip}\hline
\end{tabular}
\end{table*}

\begin{table*}
\small 
\caption{Rank-order stability of \humantic Big Five scores, as measured by Spearman's rank correlations. Reliabilities below 0.90 highlighted in yellow. Results are discussed in Sections~\ref{sec:file_format_results}, \ref{sec:url_embedding_results}, \ref{sec:source_context_results}, \ref{sec:reproducibility_and_algorithm_time_results}, and \ref{sec:participant_time_results}.}
\label{humantic_corr_big5}
\begin{tabular}{>{\raggedright}p{0.1\textwidth} | >{\raggedright}p{0.14\textwidth} | >{\centering}p{0.02\textwidth} | >{\centering}p{0.08\textwidth} | >{\centering}p{0.1\textwidth} | >{\centering}p{0.1\textwidth} | p{0.1\textwidth} |>{\centering\arraybackslash}p{0.08\textwidth}} 
\hline\noalign{\smallskip}
Facet & Input Versions &   N &  Openness &  Conscien-tiousness &  Extra-version &  Agreeable-ness &  Emotional Stability \\
\noalign{\smallskip}\hline\noalign{\smallskip}
      File Format &          De-Identified Resume (HRi1) vs. DOCX Resume (HRd1) &  89 &    0.9891 &             0.9936 &        0.9939 &         0.9927 &               0.9816 \\
    URL Embedding &  URL-Embedded Resume (HRu1) vs. De-Identified Resume (HRi1) &  86 &   \signifmedhigh 0.3988 &           \signifmedhigh  0.3845 &      \signifmedhigh  0.0772 &      \signifmedhigh   0.4190 &            \signifmedhigh   0.4040 \\
    URL Embedding &     URL-Embedded Resume (HRu1) vs. LinkedIn (HL1) &  83 &  \signifmed  0.6381 &          \signifmed   0.5470 &        \signifmed 0.5786 &      \signifmed   0.6839 &            \signifmed   0.7018 \\
   Source Context &            De-Identified Resume (HRi1) vs. LinkedIn  (HL1) &  84 &   \signifmedhigh 0.2180 &           \signifmedhigh  0.1558 &     \signifmedhigh   0.1198 &       \signifmedhigh  0.2020 &            \signifmedhigh   0.2186 \\
   Source Context &                  Original Resume (HRo1) vs. LinkedIn (HL1) &  84 &  \signifmed  0.5985 &          \signifmed   0.7124 &        \signifmed 0.5827 &       \signifmed  0.6136 &           \signifmed    0.5990 \\
   Source Context &                   Original Resume (HRo1) vs. Twitter (HT1) &  20 &  \signifhigh -0.1768 &          \signifmedhigh   0.2324 &       \signifhigh -0.1128 &      \signifhigh  -0.2316 &              \signifmedhigh 0.0692 \\
   Source Context &                          LinkedIn (HL1) vs. Twitter (HT1) &  18 &   \signifhigh -0.2158 &           \signifmedhigh  0.0000 &       \signifhigh -0.1559 &      \signifhigh  -0.1517 &           \signifhigh   -0.1125 \\
   Immediate Reproduc-ibility &            De-Identified Resume back-to-back (HRi2 vs. HRi3)  &  89 &    1.0000 &             1.0000 &        1.0000 &         0.9999 &               1.0000 \\
   Algorithm-Time &            De-Identified Resume 31 days apart (HRi1 vs. HRi2) &  89 &    0.9954 &             0.9969 &        0.9618 &         0.9921 &               0.9854 \\
 Participant-Time &                     LinkedIn 7-9 months apart (HL1 vs. HL2) &  88 &  \signifmed  0.6879 &          \signifmed   0.6928 &        \signifmed 0.7301 &       \signifmed  0.7518 &            \signifmed   0.7678 \\
 Participant-Time &                      Twitter 7-9 months apart (HT1 vs. HT2) &  21 &    1.0000 &             1.0000 &        1.0000 &         1.0000 &               1.0000 \\
 \noalign{\smallskip}\hline
\end{tabular}
\end{table*}

\begin{table*}
\small 
\caption{Significance in locational instability of \crystal DiSC scores, as measured by two-tailed Wilcoxon signed-rank test p-values. The absence of yellow highlighting indicates that all values are below both the Benjamini-Hochberg and Bonferroni-corrected thresholds based on $\alpha$ of 0.05. ``N/A'' values reflect experiments where there was zero change across the facet. Results are discussed in Sections~\ref{sec:file_format_results}, \ref{sec:url_embedding_results}, \ref{sec:source_context_results}, \ref{sec:reproducibility_and_algorithm_time_results}, and \ref{sec:participant_time_results}.}
\label{crystal_wc_disc}
\begin{tabular}{>{\raggedright}p{0.14\textwidth} | >{\raggedright}p{0.2\textwidth} | >{\centering}p{0.02\textwidth} | >{\centering}p{0.1\textwidth} | >{\centering}p{0.08\textwidth} | >{\centering}p{0.09\textwidth} | >{\centering\arraybackslash}p{0.12\textwidth}} 
\hline\noalign{\smallskip}
Facet & Input Versions &   N &  Dominance &  Influence & Steadiness &  Conscientious-ness \\
\noalign{\smallskip}\hline\noalign{\smallskip}
File Format &  Raw Text Resume (CRr1) vs. PDF Resume (CRp1) &  89 &     0.5026 &     0.4208 &      0.0173 &           0.0370 \\
   Source Context &         PDF Resume (CRp1) vs. LinkedIn (CL1) &  86 &  0.4190 &  0.0012 &      0.7010 &          0.8421 \\
   Immediate Reproducibility &    Raw Text Resume back-to-back (CRr2 vs. CRr3) &  89 &          N/A &          N/A &           N/A &                N/A \\
   Algorithm-Time &   Raw Text Resume 31 days apart (CRr1 vs. CRr2) &  89 &          N/A &          N/A &           N/A &                N/A \\
 Participant-Time &      LinkedIn 8-10 months apart (CL1 vs. CL2) &  89 &     0.7299 &     0.6518 &      0.3305 &           0.2870 \\
 \noalign{\smallskip}\hline
\end{tabular}
\end{table*}

\begin{table*}
\small 
\caption{Significance in locational instability of \humantic DiSC scores, as measured by two-tailed Wilcoxon signed-rank test p-values. Yellow highlighting indicates value below Bonferroni-corrected threshold based on $\alpha$ of 0.05. Lighter yellow indicates p-value below Benjamini-Hochberg corrected threshold and above Bonferroni-corrected threshold. ``N/A'' values reflect experiments where there was zero change across the facet. Results are discussed in Sections~\ref{sec:file_format_results}, \ref{sec:url_embedding_results}, \ref{sec:source_context_results}, \ref{sec:reproducibility_and_algorithm_time_results}, and \ref{sec:participant_time_results}.}
\label{humantic_wc_disc}
\begin{tabular}{>{\raggedright}p{0.14\textwidth} | >{\raggedright}p{0.2\textwidth} | >{\centering}p{0.02\textwidth} | >{\centering}p{0.1\textwidth} | >{\centering}p{0.08\textwidth} | >{\centering}p{0.09\textwidth} | >{\centering\arraybackslash}p{0.13\textwidth}} 
\hline\noalign{\smallskip}
Facet & Input Versions &   N &  Dominance &  Influence &  Steadiness &  Calculativeness \\
\noalign{\smallskip}\hline\noalign{\smallskip}
      File Format &          De-Identified Resume (HRi1) vs. DOCX Resume (HRd1) &  89 &     0.2510 &     0.2940 &      0.4574 &           0.2539 \\
    URL Embedding &  URL-Embedded Resume (HRu1) vs. De-Identified Resume (HRi1) &  86 &     \signifhigh 0.0000 &     0.3194 &     \signiflow 0.0005 &         \signiflow  0.0047 \\
    URL Embedding &      URL-Embedded Resume (HRu1) vs. LinkedIn (HL1) &  83 &  \signiflow   0.0066 &     0.1825 &      0.5324 &           0.1213 \\
   Source Context &            De-Identified Resume (HRi1) vs. LinkedIn  (HL1) &  84 &     \signifhigh 0.0000 &     0.0580 &   \signiflow   0.0013 &           0.3259 \\
   Source Context &                  Original Resume (HRo1) vs. LinkedIn (HL1) &  84 &     \signifhigh 0.0000 &  \signiflow  0.0050 &      0.2299 &           0.5911 \\
   Source Context &                   Original Resume (HRo1) vs. Twitter (HT1) &  20 &     0.5706 &     0.3118 &      0.1975 &           0.6874 \\
   Source Context &                          LinkedIn (HL1) vs. Twitter (HT1) &  18 &     0.0342 &     0.3247 &      0.6095 &           0.5539 \\
   Immediate Reproducibility &            De-Identified Resume back-to-back (HRi2 vs. HRi3) &  89 &     0.3173 &     0.3173 &           N/A &                N/A \\
   Algorithm-Time &            De-Identified Resume 31 days apart (HRi1 vs. HRi2) &  89 &     0.1416 &     0.5971 &      0.5690 &           0.0307 \\
 Participant-Time &                     LinkedIn 7-9 months apart (HL1 vs. HL2) &  88 &     0.0709 &     0.0800 &      0.3457 &           0.2969 \\
 Participant-Time &                      Twitter 7-9 months apart (HT1 vs. HT2) &  21 &          N/A &          N/A &           N/A &                N/A \\
 \noalign{\smallskip}\hline
\end{tabular}
\end{table*}

\begin{table*}
\small 
\caption{Significance in locational instability of \humantic Big Five scores, as measured by two-tailed Wilcoxon signed-rank test p-values. Yellow highlighting indicates value below Bonferroni-corrected threshold based on $\alpha$ of 0.05. Lighter yellow indicates p-value below Benjamini-Hochberg corrected threshold and above Bonferroni-corrected threshold. ``N/A'' values reflect experiments where there was zero change across the facet. Results are discussed in Sections~\ref{sec:file_format_results}, \ref{sec:url_embedding_results}, \ref{sec:source_context_results}, \ref{sec:reproducibility_and_algorithm_time_results}, and \ref{sec:participant_time_results}.}
\label{humantic_wc_big5}
\begin{tabular}{>{\raggedright}p{0.1\textwidth} | >{\raggedright}p{0.14\textwidth} | >{\centering}p{0.02\textwidth} | >{\centering}p{0.08\textwidth} | >{\centering}p{0.1\textwidth} | >{\centering}p{0.1\textwidth} | >{\centering}p{0.1\textwidth} | >{\centering\arraybackslash}p{0.08\textwidth}}
\hline\noalign{\smallskip}
Facet & Input Versions &   N &  Openness &  Conscien-tiousness &  Extra-version &  Agreeable-ness &  Emotional Stability \\
\noalign{\smallskip}\hline\noalign{\smallskip}
     File Format &          De-Identified Resume (HRi1) vs. DOCX Resume (HRd1) &  89 &    0.7193 &             0.9248 &        0.5306 &         0.3003 &               0.9771 \\
    URL Embedding &  URL-Embedded Resume (HRu1)  vs. De-Identified Resume (HRi1) &  86 &  \signiflow  0.0025 &             \signifhigh 0.0000 &       \signifhigh 0.0000 &         \signiflow 0.0002 &               0.2214 \\
    URL Embedding &    URL-Embedded Resume (HRu1) vs. LinkedIn (HL1) &  83 &    0.7352 &            \signifhigh 0.0000 &        0.3603 &     \signiflow   0.0068 &               0.7167 \\
   Source Context &            De-Identified Resume (HRi1) vs. LinkedIn (HL1) &  84 &  \signiflow  0.0077 &             0.3997 &       \signifhigh 0.0000 &         0.1730 &               0.6718 \\
   
   Source Context &                  Original Resume (HRo1) vs. LinkedIn (HL1) &  84 &    0.5300 &             \signiflow 0.0003 &      \signifhigh  0.0001 &         0.0221 &               0.4553 \\
   Source Context &                   Original Resume (HRo1) vs. Twitter (HT1) &  20 &   0.0121 &             0.0826 &        0.8983 &         \signiflow 0.0020 &            \signiflow   0.0010 \\
   Source Context &                          LinkedIn (HL1) vs. Twitter (HT1) &  18 &  \signiflow  0.0023 &          \signiflow   0.0047 &     \signiflow   0.0007 &       \signiflow  0.0047 &            \signiflow   0.0007 \\
   Immediate Reproduc-ibility &            De-Identified Resume back-to-back  (HRi2 vs. HRi3) &  89 &    0.1797 &             0.3173 &        0.3173 &         0.6547 &               0.6547 \\
   Algorithm-Time &            De-Identified Resume 31 days apart (HRi1 vs. HRi2) &  89 &  \signiflow  0.0071 &             0.5314 &        0.2540 &         0.0516 &               0.2424 \\
 Participant-Time &                     LinkedIn 7-9 months apart (HL1 vs. HL2) &  88 &    0.6487 &             \signifhigh 0.0000 &        0.9615 &        \signiflow  0.0072 &               0.6011 \\
 Participant-Time &                      Twitter 7-9 months apart (HT1 vs. HT2) &  21 &         N/A &                  N/A &             N/A &              N/A &                    N/A \\
 \noalign{\smallskip}\hline
\end{tabular}
\end{table*}

\end{document}